# Spacetime G-structures II: geometry of the ground states

D.H. Delphenich[†]
Physics Department
University of Wisconsin – River Falls
River Falls, WI 54022

*Abstract. This article is a continuation of a previous work that dealt with the topological obstructions to the reductions of the bundle of linear frames on a spacetime manifold for a particular chain of subgroups of GL(4). In this article, the corresponding geometrical information, such as connections, torsion, curvature, and automorphisms of the reductions will be discussed. The details are elaborated upon for a certain sequence of reductions of GL(M) when M is a four-dimensional spacetime manifold.*

**0. Introduction.** In a Part I of this work [**1**], the notion of a $G$-structure was applied to the case of the spacetime manifold and the topological obstructions to the sequence of bundle reductions that were associated with the following chain of subgroups in the affine group $A(4)$:

$$(0.1)$$

$$A(4) \leftarrow GL(4) \leftarrow GL^+(4) \leftarrow SL(4) \leftarrow SO(3,1) \leftarrow SO_0(3,1) \leftarrow SO(3) \leftarrow SO(2) \leftarrow \mathbb{Z}_2 \leftarrow \{e\}$$

were explicitly computed.

*a. Spacetime as an ordered medium.* The physical interpretation of these topological obstructions was then presented in the language of topological defects in ordered media ([1]):

*i*) A subgroup reduction $G \leftarrow H$ was considered to be a form of *spontaneous symmetry breaking*.

*ii*) The subgroup reductions for which $G/H$ was non-contractible were regarded as the *(stable) phases* of the *spacetime vacuum manifold*, which was defined to be the homogeneous space $G/H$.

*iii*) A *phase transition* in the vacuum manifold was associated with a reduction for which the homotopy type of $G/H$ changed.

*iv*) A subgroup reduction that produced a contractible $G/H$ was shown to be associated with a deformation retraction of $G$ to $H$; these reductions involved no phase transition in the vacuum manifold. Such reductions were regarded as the *(unstable) deformations* of the vacuum manifold for that phase.

*v*) The *order parameter* for a given phase was represented as either a $G$-equivariant map from the $G$-structure $G(M)$ to $G/H$ or a section of the associated homogeneous fiber bundle $G/H(M)$.

---

[†] david.delphenich@uwrf.edu

[1] The terminology presented in this article is a slight refinement of the terminology that was suggested in Part I, at least as far as the distinction between phases and states is concerned. In this Part, we shall treat "phase" as a topological concept and "state" as a geometrical one. Indeed, in a thermodynamical "phase diagram" the points are states and the connected components are the phases.

*vi*)  When $G/H$ has non-vanishing homotopy in some dimension $k$, there was a non-vanishing obstruction cocycle in dimension $k$-1 associated with the reduction, which we showed was essentially a generalized form of *topological defect* of codimension $k$.

Particular emphasis was devoted to the obstructions to defining an $SO(2)$-structure on spacetime, since previous work of the author [**2**] has suggested that such a structure is not only essential for the study of wave motion in manifolds in general, but also relates to the geometrical interpretation for the Madelung potential [**3**] that arises in the so-called "hydrodynamical" intepretation of the Klein-Gordon equation.

In the present work, the geometrical considerations that are associated with the same chain of subgroups will be examined.  In particular, we will start with an affine connection on the bundle $A(M)$ of affine frames to the spacetime manifold and examine the issues that are associated with its reducibility at each step in the sequence, and the nature of the deformation of the connection that is defined when the connection is not reducible.

Although the reduction of a subgroup by way of deformation retraction – such as $GL^+(n)$ to $SL(n)$ – has limited *topological* interest, it is, in a complementary way, of considerable *geometrical* interest.  In particular, it seems to be the case that the subgroup reductions that produced a change in the homotopy type of the vacuum manifold produced no change in the Lie algebra of the Lie subgroup in question, and vice versa. Indeed, the notion of a subgroup $H$ being reducible in $G$ will play an important role, as did the notion of deformation retract in the previous part of this article.

In this part of our research into the nature of spacetime $G$-structures and the phases of the spacetime vacuum manifold, we shall concern ourselves with these complementary geometrical notions that apply between the phase transitions.  We shall approach the geometrical issues in two parts: First, we shall examine the geometry of each of the $G$-structures in the geometrical subsequence of (0.1) and the nature of their automorphisms. Then, we shall examine the way that the irreducible part of an irreducible connection for reduction from a $G$-structure to an $H$-structure relates to a type of deformation of the underlying $H$-structure that is especially tractable to work with in geometry.  However, the study of deformations shall define Part III of this series.

So far, our latter association of $G$-structure concepts to corresponding concepts in the physics of ordered media has left out any notion of the *state* of the spacetime vacuum manifold in a given phase.  To be physically consistent in our terminology, we really need to deal with the energetics of the geometrical objects that we will be dealing with in order to justify the notion of a "ground" state (the words "equilibrium," "undeformed," and "natural" might also be used) of a phase versus the "excited" states.

However, we shall defer that effort to subsequent research and axiomatically characterize the ground state as some subspace of the affine space of all $G$-connections on a given $G$-structure that will eventually represent a "least-action" field configuration, à la gauge field theory. In this article we shall generally think of the subspace of torsionless g-connections on a $G$-structure as representing this ground state of the vacuum manifold in the phase associated with $G(M)$ and the role of torsion as being ascribed to a translational deformation (*dislocation*) of the ground state.

For the orthogonal subgroups of $GL(4)$, this subspace of torsionless metric connections will be a unique point – viz., the Levi-Civita connection – but for $GL(4)$, $GL+(4)$ and $SL(4)$ there is a 40-dimensional subspace of the affine space of respective connections that all have zero torsion.

*b. G-structures.* Now, let us summarize the basic definitions relating to *G*-structures that were introduced in [**1**] as they relate to the present discussion:

Let *M* be an *n*-dimensional differentiable manifold and let us denote its bundle of linear frames by *GL*(*M*); this bundle is a principal *GL*(*n*)-bundle over *M*. If *G* is a subgroup of *GL*(*n*) then one calls any reduction of *GL*(*M*) to a *G*-principal bundle, which we generally denote by *G*(*M*), a *G-structure* on *M* [**4-8**]. Here, an *H-reduction* of a *G*-principal bundle *P* is a submanifold of *P* that defines an *H*-principal bundle over *M* along with the restriction of the projection of *GL*(*M*) on *M*.

There are many important examples of *G*-structures in differential geometry. For example, an orientation of the manifold *M* is associated with a reduction of the linear frames to oriented linear frames, hence, a reduction from *GL*(*M*) to $GL^+(M)$. A Riemannian metric on *M* is equivalent to a reduction of *GL*(*M*) to *O*(*M*), that is reducing from linear frames to orthonormal frames. Naturally, if one has made this reduction in two steps − *GL*(*M*) to $GL^+(M)$ to *O*(*M*) − one can further reduce to *SO*(*M*), i.e., oriented orthonormal frames. Similarly, a Lorentz structure on M is a reduction from *GL*(*M*) to *O*(1,3)(*M*), i.e., Lorentz orthonormal frames. Ultimately, one might reduce to an {*e*}-structure, which is a unique choice of frame at each point of *M*, i.e., a global frame field on *M*. As this example clearly shows, reductions to subgroups are not always possible and depend upon the topology of *M*. Some other geometric structures on manifolds that can be represented by *G*-structures are exterior differential systems, symplectic structures, and almost-complex structures.

The existence of a reduction of a *G*-principle bundle $P \to M$ to an *H*-principle bundle is equivalent to the existence of a global section of the *associated homogeneous bundle* that is defined by *P* and *G/H*. This is a fiber bundle over *M* with fiber *G/H* and structure group *G* that one obtains by looking at the orbits of the action of *G* on *P×G/H* by way of $g(p, v) = (pg^{-1}, gv)$, and one denotes this bundle by $P×_G G/H$, or, when no confusion will arise, *G/H*(*M*). More directly, one can consider the action of *H* on *GL*(*M*) and think of a point of a fiber of this bundle at any $x \in M$ as representing an orbit of the action of *H* on $GL_x(M)$.

These sections, in turn, are in one-to-one correspondence with the *G*-equivariant maps from *P* to *G/H*. Since the principal bundles that we are dealing with will be frame bundles over differentiable manifolds and the homogeneous spaces *G/H* will be equivalence classes of real invertible *n×n* matrices, there will be instances when these equivariant maps take the form of tensor fields on *M*, which we call the *fundamental tensor field* of the *G*-structure that we have defined. In all cases that we will consider in the present study, these tensor fields will be functions, vector fields, and second rank tensor fields on *M*.

*c. The geometrical subsequence.* In the present work, it will not be necessary to discuss each of the steps in the aforementioned chain specifically, since the only geometrically interesting steps are the ones that involve an actual change in the Lie algebra of infinitesimal transformations. Hence, the steps that were involved with choosing orientations may have contained useful topological information, but will no longer concern us here. We can subdivide the sequence (0.1) into two subsequences, one of which we will call the *topological subsequence*:

$$GL(4) \leftarrow GL^+(4) \leftarrow SO_0(3,1) \leftarrow SO(2) \leftarrow Z_2 \leftarrow \{e\} \qquad (0.2)$$

which gives the consecutive homogeneous spaces:

$$Z_2, R^{10} \mathfrak{Z} Z_2, R^3 \mathfrak{Z} S^2, RP^2, Z_2, \qquad (0.3)$$

and the other of which we will we will call the *geometrical subsequence*:

$$A(4) \leftarrow GL^+(4) \leftarrow SL(4) \leftarrow SO_0(3,1) \leftarrow SO(3) \leftarrow SO(2) \leftarrow \{e\}. \qquad (0.4)$$

which gives the sequence of the homogeneous spaces:

$$R^4, R^+, R^9, R, S^2, S^1. \qquad (0.5)$$

In Part I, our concern was with topological obstructions to reduction, so the emphasis was on the topological sequence, since the homotopy groups of associated sequence of homogeneous spaces contained the basic elements of the obstruction cocycles.

In this part, we shall be more concerned with the geometrical subsequence. The first four homogeneous spaces that it defines are the result of deformation retractions of subgroups and are thus not associated with any change in homotopy type of the spacetime vacuum manifold. However, these deformation retractions still play a geometric role in the context of deformations of the *G*-structures and *G*-connections. The last two homogeneous spaces in the sequence are not contractible, but they nevertheless define fundamental tensor fields, in the form of spacelike unit vector fields, or spacelike unit 1-forms on the *G*-structure with values in $R^2$ and $R$, respectively. Hence, they are still of geometrical interest.

The geometrical sequence also involves only connected components of the identity, so it defines the sequence of Lie subalgebra inclusions:

$$A(4) \leftarrow gl(4) \leftarrow sl(4) \leftarrow so(3,1) \leftarrow so(3) \leftarrow R \leftarrow 0. \qquad (0.6)$$

As mentioned above, a particularly crucial role will be played in what follows by pairs of Lie algebras $(g, h)$ for which $h$ is *reductive* in $g$, i.e., $g$ is a subalgebra of $h$ with a complementary subspace $m$, so $g = h \oplus m$, that is invariant under the adjoint action of $h$, i.e., $[h, m] \subset m$. The complementary subspace $m$ then takes on the role of infinitesimal deformations of the transformations in $h$. As we shall observe, all of the successive pairs in the aforementioned sequence, after the first one, are reductive.

The primary focus in the present work is on starting with a chain of subgroups of $GL(4)$ that go all the way down to $\{e\}$ and examining the geometrical data that are associated with each corresponding reduction of $GL(M)$ when $M$ is the spacetime manifold; actually, we shall start a step higher with the bundle of *affine* frames, so we can also examine the reduction to the bundle of linear frames, as well. For the sake of specificity, we will consider spacetime to be a connected Haussdorff separable four-dimensional manifold without boundary. However, we shall regard its Lorentz structure only in the context of the other reductions of $GL(M)$.

The example of reducing from the affine group to the general linear group shows that a subgroup $H$ can be a deformation retract of a group $G$ and still not be reducible in $G$. However, the case of a Lie subgroup $H$ that is a deformation retract of a Lie group $G$ and is reducible in $G$ is particularly useful as a bridge between topology and geometry. Nevertheless, we shall defer the discussion of how the topology in Part I relates to the geometry in Part II until Part IV, which will examine the issue of integrability.

**1. Reduction of linear connections.** For a given $n$-dimensional differentiable manifold $M$ the geometrical structure of its bundle of linear frames $GL(M)$ is largely contained in two 1-forms on $GL(M)$: the canonical 1-form $\theta^i$ on $GL(M)$ with values in $\mathbf{R}^n$, which is associated with the bundle $GL(M)$ itself, and the 1-form $\omega_j^i$ with values in $\mathfrak{gl}(n)$ that is associated with a choice of linear connection on $M$, as well as the exterior covariant derivatives of these two 1-forms, in the form of the torsion and curvature 2-forms, respectively. These equations for the exterior covariant derivatives are called the *Cartan structure equations*. A second exterior covariant differentiation produces the *Bianchi identities*.

If one reduces from $GL(M)$ to a $G$-structure $G(M)$ that is defined by a fundamental tensor field $t: GL(M) \rightarrow GL(4)/G$ then one has another geometrical object to account for in the structure equations and Bianchi identities. One must also account for the nature of the canonical 1-form $\theta^i$ and a g-connection $\varpi$ on $G(M)$ that one obtains by reducing from the corresponding objects on $GL(M)$. We shall therefore treat the geometry of $G(M)$ as if it were defined by $\{t, \theta^i, \varpi\}$, and its first two exterior covariant derivatives.

We shall now go over the preceding remarks in detail.

*a. Structure equations for linear connections.* The canonical 1-form $\theta^i$ on $GL(M)$ is defined by:

$$\theta^i(\bar{\mathbf{v}}) = v^i, \tag{1.1}$$

when $\bar{\mathbf{v}} \in T_{\mathbf{e}}(GL(M))$ and $v^i \mathbf{e}_i = \pi_*(\bar{\mathbf{v}})$, and $\pi: GL(M) \rightarrow M$ is the bundle projection.

A *linear connection* on $GL(M)$ is either a smooth $GL(n)$-invariant sub-bundle of $T(GL(M))$ that is complementary to the vertical sub-bundle:

$$T(GL(M)) = H(GL(M)) \oplus V(GL(M)), \tag{1.2}$$

i.e., a horizontal sub-bundle of $T(GL(M))$, or a smooth $\mathrm{Ad}^{-1}$-equivariant 1-form $\omega_j^i$ on $GL(M)$ with values in $\mathfrak{gl}(n)$. The relation between them is that the horizontal sub-bundle is the bundle of annihilating subspaces of $\omega_j^i$, i.e., the restriction of $\omega_j^i$ to $H_{\mathbf{e}}(GL(M))$ is zero for each $\mathbf{e} \in GL(M)$. If one *defines* the torsion and curvature 2-forms that are associated with $\omega_j^i$ as the horizontal parts of $d\theta^i$ and $d\omega_j^i$:

$$\Theta^i(X, Y) = d\theta^i(H(X), H(Y)) \tag{1.3a}$$

$$\Omega(X, Y) = d\omega(H(X), H(Y)), \tag{1.3b}$$

then one can also describe the geometry of $GL(M)$ by the *Cartan structure equations:*

$$\Theta^i = \nabla\theta^i = d\theta^i + \omega^i_j \wedge \theta^i, \tag{1.4a}$$

$$\Omega = \nabla\omega = d\omega + \omega \wedge \omega. \tag{1.4b}$$

Here, we are introducing the abbreviations $\omega^i_j \wedge \theta^i$ and $\omega \wedge \omega$  to mean:

$$(\omega^i_j \wedge \theta^i)(X,\,Y) = \tfrac{1}{2}\{\,\omega^i_j(X)\theta^i(Y) - \omega^i_j(Y)\theta^i(X)\} \tag{1.5a}$$

$$(\omega \wedge \omega)(X,\,Y)\quad = \tfrac{1}{2}[\omega(X),\,\omega(Y)], \tag{1.5b}$$

and, more generally:

$$(\omega_1 \wedge \omega_2)(X,\,Y) = \tfrac{1}{4}\{[\omega_1(X),\,\omega_2(Y)] - [\omega_1(Y),\,\omega_2(X)]\}$$

$$= (\omega_2 \wedge \omega_1)(X,\,Y). \tag{1.5c}$$

One refers to the operator $\nabla$ that is defined in either case as the *exterior covariant differential.*  A further application of that operator to the structure equations gives the *Bianchi identities:*

$$\nabla\Theta^i = \nabla^2\theta^i = \Omega^i_j \wedge \theta^i \tag{1.6a}$$

$$\nabla\Omega = \nabla^2\omega = 0. \tag{1.6b}$$

Note that in general the exterior covariant differential operator differs from the exterior derivative operator in that it is not always nilpotent; for a $k$-form $\alpha$ that takes its values in a vector space upon which the Lie algebra $\mathrm{gl}(n)$ acts linearly, one has:

$$\nabla\alpha = d\alpha + \omega \wedge \alpha, \qquad \nabla^2\alpha = \Omega \wedge \alpha. \tag{1.7}$$

The specific form that the expression $\omega \wedge \alpha$ takes depends upon the vector space in which $\alpha$ takes its values and the nature of the action of $\mathrm{gl}(n)$ upon it.

Something that is generally overlooked in the Bianchi identities is that when one is dealing with connections with non-vanishing torsion, there are further non-vanishing exterior covariant derivatives to contend with:

$$\nabla^3\theta^i = \nabla^2\Theta^i = \Omega^i_j \wedge \Theta^j \tag{1.8a}$$

$$\nabla^4\theta^i = \nabla^3\Theta^i = \Omega^i_j \wedge \Omega^j_k \wedge \theta^k, \tag{1.8b}$$

etc.  Of course, if one assumes that the manifold in question is four-dimensional then any $k$-form for $k > 4$ will vanish.  This situation with the existence of higher derivatives of $\theta^i$ is closely related to the issue of the integrability of the $H(GL(M))$ as a differential system on $GL(M)$, which, in turn, relates to the fact that the Lie algebra $\mathrm{gl}(n)$ has prolongations of all orders, which we shall discuss later.

*b.  Augmented structure equations for reduced connections.*  When one reduces from $GL(M)$ to a $G$-structure on $M$, $G(M)$, the geometry of $G(M)$ will now be defined by a canonical 1-form $\theta^i$, a connection form $\varpi$, and, in some cases, a fundamental tensor field $t$; in this discussion, we shall assume that such a $t$ is defined by the reduction to $G(M)$.

Note that since $t{:}G(M) \to GL(4)/G$, and the homogeneous space in which $t$ takes its values is generally represented by equivalence classes of $4{\times}4$ real matrices, the tensor field on $M$ that $t$ defines (directly, anyway) will be of rank at most two.

The canonical 1-form on $G(M)$ is simply the restriction of $\theta^i$ to $G(M)$, but the situation that is associated with reducing the connection form $\omega^i_j$ is somewhat more involved.  In particular, the reduced connection must take its values in g, the Lie algebra of $G$, and be $G$-equivariant.  The latter requirement is automatic by restriction, but the former one is not.  If the $G$-structure $G(M)$ is associated with a fundamental tensor field $t$ then a linear connection can be reduced to a $G$-connection on $G(M)$ iff:

$$Q = \nabla t = dt + \omega \wedge t = 0. \tag{1.9}$$

(In the case of a Riemannian metric this simply says that the linear connection would have to also be a metric connection.) We shall refer to the 1-form $Q$ as the *irreducibility* 1-form of $\omega$.  Although $t$ is a 0-form on $GL(M)$ with values in $GL(n)/G$, we are using the notation $\omega \wedge t$ to indicate that $\omega$ takes its values in $\mathfrak{gl}(n)$, which acts on $GL(n)/G$ by way of exponentiating the action of $GL(n)$ on $GL(n)/G$.

A further covariant differentiation gives:

$$\nabla Q = \nabla^2 t = \Omega \wedge t, \tag{1.10}$$

which must naturally vanish for a reducible connection.

If a linear connection $\omega$ is reducible to a $G$-connection $\varpi$ then we summarize the foregoing discussion by saying that we are simply augmenting our geometry by the addition of $t$ and its first two exterior covariant derivatives.  Hence:

*i)*      The geometry of $G(M)$ is defined by $\{t, \theta^i, \varpi\}$.

*ii)*     The augmented Cartan structure equations for $\varpi$ are:

$$Q = \nabla t = d\varpi + \varpi \wedge t = 0, \tag{1.11a}$$

$$\Theta^i = \nabla \theta^i = d\theta^i + \varpi \wedge \theta^i, \tag{1.11b}$$

$$\Omega = \nabla \varpi = d\varpi + \varpi \wedge \varpi, \tag{1.11c}$$

*iii*)    The augmented Bianchi identities that follow from these structure equations are then:

$$\nabla Q = 0, \qquad \nabla \Theta^i = \Omega \wedge \theta^i, \qquad \nabla \Omega = 0. \tag{1.12}$$

A particularly useful class of Lie algebras for the purposes of reducing connections is the class of reductive Lie algebras.  If the Lie subalgebra h is *reductive* in the Lie algebra g, i.e.:

*a)*  g = m $\oplus$ h, for some linear subspace m $\subset$ g

*b)*  [h, m ] $\subset$ m,

then one can decompose the $G$-connection $\omega^i_j$ into a 1-form $\tilde{\omega}^i_j$ with values in $\mathfrak{h}$ and a 1-form $\tau^i_j$ on $G(M)$ with values in $\mathfrak{m}$:

$$\omega^i_j = \tilde{\omega}^i_j + \tau^i_j \tag{1.13}$$

in such a way that the restriction of $\tilde{\omega}^i_j$ to $H(M)$ is actually a reduced $H$-connection; in particular $\tilde{\nabla}\, \tilde{t} = 0$.

The equivariant 1-form $\tau$ also measures the irreducibility of $\omega$ under the reduction in question, as does $Q$. Indeed, if $\nabla$ represents the exterior covariant derivative that is defined by $\omega$ then the fact that the connection $\varpi$ must satisfy $\tilde{\nabla}\, t = 0$ gives an algebraic relation between $\tau$ and $Q$:

$$\tau \wedge t = Q \tag{1.14}$$

that one can generally solve for $\tau$ when one is given $Q$, according to the nature of $t$.

If $\omega$ is chosen from the outset then we can think of $\tau$ as being *induced* by the reduction from $G(M)$ to $H(M)$. Note that if one is given $\tau$ then one cannot generally define a *unique* reduction from $G(M)$ to $H(M)$ that induces it; in particular, $\tau$ can be 0 for more than one reduction since there can be more than one reduction for which $\omega$ is reducible.

We shall return to examining the role of $\tau$ in greater detail in the Part III of this series, which will be concerned with the deformation of the geometrical structures that we discuss in this Part.

   *c. Parallel translation and geodesics for G-structures.* Note that $\theta^i$ maps each horizontal subspace on $G(M)$ isomorphically to $\mathbf{R}^n$; in particular, if $\hat{\mathbf{v}} \in T_{\mathbf{e}}G(M)$ then:

$$\theta^i(\hat{\mathbf{v}}) \quad = v^i, \tag{1.15}$$

where $\pi_* \hat{\mathbf{v}} = v^i \mathbf{e}_i$. Conversely, the inverse isomorphism defined by $\theta^i$ defines a lift of any tangent vector field on $M$ to a horizontal tangent vector field on $G(M)$. More precisely, if $\mathbf{v} \in T_x(M)$ is a vector tangent to $M$ and $\mathbf{e}_i$ is a $G$-frame at $x$, so the components of $\mathbf{v}$ are $v^i$, then one can define the horizontal lift of $\mathbf{v}$ to $\mathbf{e}$ as:

$$\hat{\mathbf{v}}(\mathbf{e}) = \theta^{-1}(v^i) = v^i E_i, \tag{1.16}$$

where the $E_i = \theta^{-1}(\delta_i)$, $i = 1, \ldots, n$ are the basic vector fields on $G(M)$. Since the isomorphism that is defined by $\theta^i$ at $\mathbf{e}$ is invertible only when one restricts its range to a choice of horizontal subspace, this lift is dependent upon the choice of connection on $G(M)$.

   One can further extend $\hat{\mathbf{v}}$ to a $G$-invariant horizontal vector field on the fiber $G_x(M)$:

$$\hat{\mathbf{v}}(\mathbf{e}g) = \theta^{-1}(g^{-1}v^i). \tag{1.17}$$

The existence of such a horizontal lift of tangent vectors allows one to define the parallel translation of $G$-frames along curves in $M$. If $\gamma(\tau)$ is a curve in $M$ and $\mathbf{v}(\tau)$ is its velocity vector field then there is a unique horizontal lift of $\mathbf{v}(\tau)$ to a $G$-invariant vector field $\hat{\mathbf{v}}(\tau)$ on the fibers of $G(M)$ over $\gamma(\tau)$. If $\mathbf{e}_i$ is a $G$-frame at $\gamma(0)$ then, at least for some neighborhood $(-\varepsilon, +\varepsilon)$ of 0, one can find a unique integral curve $\hat{\gamma}(\tau)$ of the vector field $\hat{\mathbf{v}}(\tau)$ that passes through $\mathbf{e}_i$. The frames $\mathbf{e}_i(\tau)$ that $\hat{\gamma}(\tau)$ passes through for subsequent $\tau$ are then referred to as *parallel translates* of the initial frame. The frame field $\mathbf{e}_i(\tau)$ along $\gamma$ satisfies the equation:

$$0 = \frac{d\mathbf{e}_i}{d\tau} + \omega_i^{\ j}(\hat{\mathbf{v}})\mathbf{e}_j. \tag{1.18}$$

Since any tensor field $T$ on $M$ that transforms as a representation of $G$ can be described by its components $T_{\cdots}^{\cdots}$ with respect to a given $G$-frame at each point of $M$ one sees that having a definition of the parallel translation of frames then gives one a definition of parallel translation for a more general tensor field. In particular, $T$ is *parallel* along $\gamma$ if its components with respect to any parallel $G$-frame field along $\gamma$ are constant. When one considers the possibility that $\mathbf{v}$ itself is parallel translated along $\gamma$, one arrives at the concept of a *geodesic* for the chosen connection; i.e., $\gamma$ is a geodesic iff the components $v^i$ of $\mathbf{v}$ are constant along $\gamma$ for any parallel frame field along $\gamma$. This gives the equation:

$$0 = \nabla_{\mathbf{v}}\mathbf{v} = \left( \frac{dv^i}{d\tau} + \omega_j^{\ i}(\mathbf{v})v^j \right)\mathbf{e}_i. \tag{1.19}$$

One can also eliminate the reference to $\gamma$ in the case that one has a local $G$-frame field on $M$; say $\mathbf{e}: U \to G(M)$. One then says that $\mathbf{e}$ is *parallel* iff $D\mathbf{e}$ takes each $T_x(U)$ to a horizontal subspace in $T_{\mathbf{e}}(G(U))$. The local frame field $\mathbf{e}$ then satisfies the equation:

$$\nabla\mathbf{e}_i = D\mathbf{e}_i + \omega_i^{\ j} \otimes \mathbf{e}_j = 0. \tag{1.20}$$

Here, we have implicitly pulled down the connection 1-form from $G(U)$ to $U$ by way of $\mathbf{e}$. Analogously, one then defines parallel tensor fields on $U$ as ones that have constant components with respect to any parallel local frame field. In particular, one can define a *geodesic vector field* this way: a vector field $\mathbf{v}$ on $M$ is geodesic for the chosen $G$-connection $\omega$ iff the components of $v^i$ are constant with respect to any local frame field $\mathbf{e}$ that is parallel for $\omega$.

2. **The space of g-connections on a $G$-structure**. The space G of g-connections on a $G$-structure $G(M)$ is an affine space that is modeled on the vector space $\Lambda_{eq}^1(G(M); \mathfrak{g})$ of $\mathrm{Ad}^{-1}$-equivariant 1-forms on $G(M)$ with values in g. This is because even though the sum of two g-connections does not have to be another g-connection, nevertheless when one is given two connections, $\omega$ and $\omega'$ one can always find an element $\tau$ of a model vector space, namely, $\Lambda^1_{eq}(G(M); \mathfrak{g})$, that acts on G in such a way that $\omega' = \omega + \tau$ is well defined. If one writes this as $\omega' - \omega = \tau$ then one sees that $\tau$, which is often referred to as the *difference 1-form*, plays the role of the displacement vector that takes $\omega$ to $\omega'$ under the action of $\Lambda^1_{eq}(G(M); \mathfrak{g})$.  However, since the 1-form on $G(M)$ that takes everything to zero does not define a connection, one cannot generally define a unique origin to G.

However, as is well known, among all of the metric connections $\omega$ on an $O(n)$-structure or $O(p, q)$-structure, there is exactly one of them – the Levi-Civita connection $\omega_0$ – whose torsion tensor vanishes.  The existence of such a uniquely defined connection would allow one to *choose* a sort of "origin" in the affine space.  For instance, one can describe the space of all metric connections as essentially deformations of the Levi-Civita connection by means of torsion, since there is a one-to-one correspondence between metric connections $\omega$ and the deformation 1-forms $\tau$ that one obtains as $\tau = \omega - \omega_0$.  More generally, when one considers the affine space of $\mathfrak{g}$-connections on a $G$-structure the question of whether the space $G_0$ of torsionless $\mathfrak{g}$-connections is empty, consists of a single point, or defines a higher-dimensional space must revert to the nature of $G$ itself.

One can find more structure in G by means of the action of the group of $G$-gauge transformations on G.  However, we shall return to that discussion after we have first discussed automorphisms of $G$-structures.

In order to examine the nature of $G_0$, one first introduces the notion of the *structure tensor* for the $G$-structure.  First, we define part of a bigraded complex on $\mathfrak{g}$, which we regard as a subalgebra of $\mathfrak{gl}(n)$, by defining:

$$C^{1,1}(\mathfrak{g}) = \mathbf{R}^n \otimes \mathfrak{g}, \qquad C^{0,2}(\mathfrak{g}) = \Lambda^2(\mathbf{R}^n) \otimes \mathbf{R}^n \qquad (2.1)$$

and a coboundary operator:

$$\partial: C^{1,1}(\mathfrak{g}) \to C^{0,2}(\mathfrak{g}), \qquad a \infty \partial a \qquad (2.2)$$

where

$$\partial a(X, Y) = a(X)Y - a(Y)X. \qquad (2.3)$$

It is important to know when $\partial$ is surjective or even bijective.  One has the result [**7**] that if $n > 3$ then if $\partial$ is surjective then $G$ must be one of the following Lie algebras ([2]):

$$\mathfrak{gl}(n), \mathfrak{sl}(n), \mathfrak{co}(p, q), \mathfrak{so}(p, q), \mathfrak{gl}(n, W(1)), \mathfrak{gl}(n, W(1); \mathbf{C}),$$

and if $n > 3$ then $\partial$ is bijective iff $G = \mathfrak{so}(p, q)$.  In either event – surjectivity or bijectivity – the space:

$$H^{0,2}(\mathfrak{g}) = C^{0,2}(\mathbf{G})/\partial C^{1,1}(\mathfrak{g})$$

vanishes.  Since an element of $C^{0,2}(\mathfrak{g})$ is an antisymmetric bilinear product on $\mathbf{R}^n$, the space $H^{0,2}(\mathfrak{g})$ is also closely related to the space of all Lie algebras on $\mathbf{R}^n$.  The latter space is not however a linear space, but an algebraic set.

---

[2]  $\mathfrak{gl}(n, W(1))$ is the Lie algebra of the subgroup of $GL(n; \mathbf{R})$ that leave a line $W(1)$ through the origin invariant; $\mathfrak{gl}(n, W(1); \mathbf{C})$ is the corresponding subgroup of $GL(n; \mathbf{C})$.  More generally, one defines $\mathfrak{gl}(n, W(k))$ and $\mathfrak{gl}(n, W(k); \mathbf{C})$ when $W(k)$ is a $k$-plane through the origin.

If we let $q$ denote the natural projection:

$$q: C^{0,2}(\mathrm{g}) \rightarrow H^{0,2}(\mathrm{g}).$$

and we represent the torsion 2-form $\Theta^i$ of a $G$-connection $\omega$ on $G(M)$ as a $G$-equivariant map:

$$\Theta: G(M) \rightarrow C^{0,2}(\mathrm{g}), \quad \mathbf{e}_i \infty \Theta^i_{jk} \tag{2.4}$$

where:

$$\Theta^i = \tfrac{1}{2}\Theta^i_{jk}\theta^j \wedge \theta^k, \tag{2.5}$$

(of course, $\theta^i$ is the canonical 1-form on $G(M)$) then we then define the (first order) *structure tensor* of $G(M)$ as:

$$c = q \bullet \Theta. \tag{2.6}$$

Note that $c = 0$ for all of the Lie algebras in our canonical sequence (0.2) except for $A(4)$ and $\{e\}$; in the case of $SO(2)$ the vanishing of $c$ follows from the vanishing of $\Lambda^2(\mathbf{R})$.

In the case of $\{e\}$, the structure tensor actually defines a generalization of the structure constants of a Lie group to the structure functions of a parallelizable manifold $M$.  If $\mathbf{e}_i$ is a global frame field on $M$ and $\theta^i$ its reciprocal coframe field then the structure functions $c^i_{jk}(x)$ are defined by:

$$[\mathbf{e}_i(x), \mathbf{e}_j(x)] = c^k_{ij}(x)\mathbf{e}_k(x) \tag{2.7a}$$

or dually:

$$d\theta^i = -c^i_{jk}\theta^j \wedge \theta^k. \tag{2.7b}$$

The structure tensor on $\{e\}(M)$ then takes $\mathbf{e}_i(x)$ to $c^i_{jk}(x)$.

Clearly, if the torsion tensor $\Theta$ vanishes then so does $c$; in fact, a partial converse obtains, as well.  Namely, there exists a torsionless g-connection on $G(M)$ iff $c = 0$.  Hence, the space $\mathrm{G}_0$ of torsionless g-connections is non-empty for all $G$ in our sequence, except $A(4)$, for which torsion is not defined, and $\{e\}$.  However, $\mathrm{G}_0$ consists of a unique connection only for the orthogonal subgroups in the sequence, namely, $SO(3, 1)$, $SO(3)$, and $SO(2)$.  In the case of $\{e\}$ − i.e., a global frame field $\mathbf{e}_i$ − in order for $c$ to be 0 the members of $\mathbf{e}_i$ would have to commute, in which case, $M$ would have to be locally diffeomorphic to $\mathbf{R}^n$; if $M$ were compact and connected, it would have to be the $n$-torus $T^n$.

When the torsionless g-connection is unique – say $\omega_0$ – then for any other g-connection $\omega \in \mathrm{G}$ the difference 1-form $(^3) \tau = \omega - \omega_0$ can be solved in terms of the torsion 2-form $\Theta^i$ of $\omega$ by using the structure equation for torsion and the fact that the torsion 2-form for $\omega_0$ is 0; that much gives:

$$\Theta^i = \tau \wedge \theta^i. \tag{2.8}$$

---

[3] Also referred to as the *contortion tensor* for the connection $\omega$.

If we express $\Theta^i$ and $\tau$ in terms of the frame $\theta^i$ as $\Theta^i = \frac{1}{2}\Theta^i_{jk}\theta^j \wedge \theta^k$ and $\tau^i_j = \tau^i_{jk}\theta^k$ then equation (2.8) gives:

$$\Theta^i_{jk} = \tau^i_{jk} - \tau^i_{kj}, \tag{2.9}$$

which is solved by symmetrization to give:

$$\tau^i_{jk} = \frac{1}{2}g^{im}(\Theta_{m:jk} + \Theta_{j:mk} + \Theta_{k:mj})\,. \tag{2.10}$$

Note that this expression would be undefined unless there were a metric tensor $g^{ij}$ associated with the reduction. Fortunately, the only subgroups for which $\omega_0$ is unique will satisfy this requirement. Equations (2.9) and (2.10) then establish a linear isomorphism between the space of all $\Theta^i$ and the space of all $\tau$.

When the torsionless $G$-connections are not unique then some difference 1-forms will take torsionless connections to other torsionless connections. In such a case, we have that $\tau$ must satisfy:

$$\tau \wedge \theta^i = 0, \tag{2.11}$$

which gives:

$$\tau^i_{jk} = \tau^i_{kj}\,. \tag{2.12}$$

Hence, the affine subspace of torsionless connections $G_0$ will be modeled on the vector space of all triply-indexed arrays of real functions $\tau^i_{jk}$ that are symmetric in their lower indices, namely $R*^n(R*^n \otimes R^n$. One can then think of the space G as being foliated by the translates of $G_0$; any two connections in the same translate will have the same torsion and differ by a 1-form $\tau \in G/G_0$ that satisfies (2.11).

## 3. Automorphisms of *G*-structures.

In keeping with our general program heretofore, we regard the essential geometrical data for a $G$-structure $G(M)$ as being derivable from the set $\{t, \theta^i, \omega\}$. We first examine the nature of the finite transformations – i.e., equivariant diffeomorphisms – of $G(M)$ that preserve these objects individually, and then we examine the infinitesimal transformations that generate them. However, since the traditional approach of differential geometry is to start with diffeomorphisms and vector fields on $M$ and then lift them to $G(M)$, we shall include the usual definitions, despite the fact that our eventual concern is more oriented towards dealing with $G(M)$ directly.

*a. Finite automorphisms.* When we say "lift" the specific definition we are using is as follows: Let $f: M \to N$ be a diffeomorphism. The map $Df$ not only defines an invertible linear map from $T_x(M)$ to $T_{f(x)}(N)$ for every $x \in M$, it also defines a diffeomorphism:

$$\hat{f}\Big|_x : GL_x(M) \to GL_{f(x)}(N), \qquad \mathbf{e}_x \propto Df\Big|_x(\mathbf{e}_x)\,. \tag{3.1}$$

This means we can also define a diffeomorphism:

$$\hat{f}: GL(M) \to GL(N), \tag{3.2}$$

which we call the *lift* of $f$ to $GL(M)$. $\hat{f}$ has the property that it is $GL(n)$-*equivariant*, i.e.:

$$\hat{f}(\mathbf{e}_x g) = \hat{f}(\mathbf{e}_x)g, \qquad \text{for all } x \in M, \mathbf{e}_x \in GL_x(M) \text{ and } g \in GL(n), \tag{3.3}$$

which derives from the fact if $\mathbf{f}_x$ is another linear frame in $T_x(M)$ then there is a (matrix) element $g \in GL(n)$ that takes $\mathbf{e}_x$ to $\mathbf{f}_x$, and another one that takes $Df(\mathbf{e}_x)$ to $Df(\mathbf{f}_x) = Df(\mathbf{e}_x g)$, which equals $Df(\mathbf{e}_x)g$, by the linearity of $Df$ on the tangent spaces. Equivariance has the effect of taking orbits to orbits, i.e., fibers of $GL(M)$ to fibers of $GL(N)$. However, as we shall see later, a map of $G$-spaces that takes orbits to orbits does not have to be equivariant.

When $G$ is a subgroup of $GL(n)$ one can always restrict the lift $\hat{f}$ of a diffeomorphism $f$ to $G(M)$. However, the issue of whether the image of $G(M)$ under $\hat{f}$ is still a $G$-structure on $N$ depends on the nature of the differential $Df|_x$ at each $x \in M$; for instance, it might not take an orthonormal frame to another orthonormal frame. Similarly, the issue of $G$-equivariance is also questionable when $G$ is a proper subgroup of $GL(M)$.

One says that a diffeomorphism $f: M \to N$ defines an *isomorphism* of a $G$-structure $G(M)$ on $M$ with a $G$-structure $G(N)$ on $N$ iff $\hat{f}$ takes a $G$-frame in $T_x(M)$ to a $G$-frame in $T_{f(x)}(N)$ for every $x \in M$. (Note that it would not accomplish anything to generalize beyond local diffeomorphisms since $Df$ would not respect the dimension of a frame unless it were of maximal rank.) An *automorphism* of a $G$-structure on $M$ is then a $G$-structure isomorphism of $M$ with itself.

Since the set of all automorphisms forms a group under composition, one also wants to know whether it also forms a Lie group, more specifically. Indeed, this is always the case when the Lie group $G$ is of finite type ([4]), or when $M$ is compact and the Lie algebra g does not contain a linear endomorphism of rank 1 (see Fujimoto [7]). However, the automorphisms of $GL(n)$-structures and $SL(n)$-structures fail these tests, so their automorphisms groups do not generally define Lie groups.

For instance, an isomorphism of an $SL(n)$-structure on $M$ is a volume-preserving diffeomorphism of $M$, an isomorphism of an $O(p,q)$-structure is an isometry of the corresponding metric or pseudo-metric, and an isomorphism of an $\{e\}$-structure simply takes the global frame field on one parallelizable $M$ to the global frame field on another diffeomorphic manifold $N$. An automorphism of an $\{e\}$-structure on $M$ is a diffeomorphism $f: M \to M$ whose differential takes the frame at $x$ to the frame at $f(x)$ for all $x \in M$; i.e., $f_*\mathbf{e} = \mathbf{e}$, where $\mathbf{e}: M \to GL(M)$ is the global frame field that defines the $\{e\}$-structure.

One could also look at $G$-equivariant automorphisms of $G(M)$ in their own right. By equivariance, one can unambiguously define the *projection* of such a map $\hat{f}: G(M) \to G(M)$ onto $M$ by $f(x) = \pi(\hat{f}(\mathbf{e}_x))$ for all $x \in M$ and any arbitrary $\mathbf{e}_x \in G_x(M)$. Here, we are using $\pi$ to denote the bundle projection $\pi\colon G(M) \to M$. One sees that the projection of a $G$-equivariant diffeomorphism of $G(M)$ is a diffeomorphism of $M$.

---

[4] Special cases of this fact were first proved for isometries of a Riemannian manifold by Myers and Steenrod, and for conformal transformations and automorphisms of $\{e\}$-structures by Kobayashi.

Along with the general case, there will be $G$-equivariant automorphisms of $G(M)$ that project to the identity map on $M$. These will take $G$-frames at a given point to other $G$-frames at the same point. By analogy with the gauge field theories in physics, we refer to $G$-equivariant automorphisms of $G(M)$ that cover the identity as *gauge transformations* of $G(M)$. The effect of equivariance in this case is to associate a unique element $g \in G$ to each $x \in M$. Hence, a gauge transformation of $G(M)$ is also represented by a smooth map:

$$\Gamma\colon M \to G, \qquad x \infty\ g. \tag{3.4}$$

In this respect, there is an advantage to dealing with frame bundles as opposed to more general principal fiber bundles, since the aforementioned statement is generally only locally well defined when the action of $G$ is on the individual fibers of the bundle, not on the bundle itself. If one represents $G$ by $n \Im n$ matrices relative to some basis on $\mathrm{R}^n$ then the matrix $A^i_j \in G$ acts on a $G$-frame $\mathbf{e}_i \in G_x(M)$ in the natural way on the right, namely $\mathbf{e}_i$ goes to $\mathbf{e}_i [A^{-1}]^i_j$.

It is natural to wonder whether an arbitrary $G$-equivariant automorphism $\tilde{f}$ of $G(M)$ can be factored into the product of the lift of a diffeomorphism $f$ on $M$ and a gauge transformation $\Gamma$, and whether this factorization is unique. One can define such a decomposition by projecting $\tilde{f}$ onto a diffeomorphism $f$ of $M$ and then lifting $f$ to another $G$-equivariant automorphism $\hat{f}$ on $G(M)$. The gauge transformation $\Gamma$ that takes $\hat{f}$ to $\tilde{f}$ is defined by:

$$\Gamma(x)\,\hat{f}(\mathbf{e}_x) = \tilde{f}(\mathbf{e}_x). \tag{3.5}$$

The uniqueness of the decomposition:

$$\tilde{f} = \Gamma \hat{f} \tag{3.6}$$

follows from the uniqueness of the lift. The fact that any two $G$-invariant diffeomorphisms on $G(M)$ that differ by such a $\Gamma$ will project to the same $f$ shows that $\Gamma$ does not have to be the identity.

When the $G$-structure $G(M)$ is defined by a fundamental tensor field $t$, one also has the result that *any* automorphism $\hat{f}$ of $G(M)$ must preserve $t$:

$$\hat{f}\,^*t = t. \tag{3.7}$$

In the case where $t$ is a metric this means that $\hat{f}$ is an isometry. Indeed, one could use (3.7) as a definition of an automorphism of $G(M)$.

Now, let us look at the effect that a $G$-invariant diffeomorphism $\hat{f}$ of $G(M)$ with itself has on $\theta^i$. Suppose $x \in M$, $\mathbf{e}_x \in G_x(M)$, and $\mathbf{v} \in T_{\mathbf{e}}(G(M))$. If $\hat{f}$ takes $\mathbf{e}_x$ to $\hat{f}(\mathbf{e}_x) = \mathbf{e}'_i$ then it also takes $\mathbf{v}$ to $\hat{f}_* \mathbf{v}$. Hence:

$$\theta^i(\hat{f}_* \mathbf{v}) = (\hat{f}^* \theta^i)(\mathbf{v}) = v'^i, \tag{3.8}$$

where $\pi_* \hat{f}_* \mathbf{v} = v'^i \mathbf{e}'_i$. If we assume that:

$$\hat{f}^* \theta^i = \theta^i \qquad (3.9)$$

then we must have that $v'^i = v^i$. Hence, such a diffeomorphism defines a *parallelism* between the $G$-frames at any $x \in M$ and the $G$-frames at $f(x)$ if we say that parallelism of tangent vectors at those two points is defined by equality of their components with respect to any choice of $\mathbf{e}_x$ and $\hat{f}(\mathbf{e}_x)$; by $G$-equivariance, equality for one pair of $G$-frames implies equality for any other. Note that such a diffeomorphism does not define a horizontal complement to the vertical sub-bundle of $T(G(M))$, in general.

Suppose $\omega$ is a $G$-connection on $G(M)$. A $G$-equivariant diffeomorphism $\hat{f}$ of $G(M)$ to itself is called an *affine transformation* of $G(M)$ iff:

$$\hat{f}^* \omega = \omega. \qquad (3.10)$$

It is clear that if a $G$-structure $G(M)$ is defined by the fundamental tensor field $t$ then a $G$-equivariant diffeomorphism that preserves $t$ will preserve any $G$-connection on $G(M)$; i.e., every automorphism of $G(M)$ gives an affine transformation. For instance, an isometry will preserve a metric connection.

Recall that defining a linear connection can be regarded as equivalent to defining a global frame field on $GL(M)$; i.e., an $\{e\}$-structure. Because affine transformations can also be regarded as a special type of automorphism for this $\{e\}$-structure, the group of all affine transformations also forms a Lie group.

*b. Gauge transformations and geometry.* Now let us assume that the automorphism $\hat{f}^*$ in question projects to the identity and observe its effects on the geometric data $\{t, \theta^i, \varpi\}$ and their exterior covariant derivatives. We represent the resulting $G$-gauge transformation by $\Gamma: M \to G$, which acts on any $\mathbf{e}_i \in G_x(M)$ by way of:

$$\mathbf{e}_i \Gamma^{-1}(x) = [\Gamma^{-1}(x)]_i^j \mathbf{e}_j. \qquad (3.11)$$

Since any automorphism must preserve $t$ we still have, *a fortiori*:

$$\hat{f}^* t = t. \qquad (3.12)$$

Since G takes a $G$-frame $\mathbf{e}_i$ at any $x \in M$ to another $G$-frame $\mathbf{e}_i \Gamma^{-1}$ at $x$, it will also take the reciprocal coframe $\theta^i$ to $\Gamma \theta^i$. Since this coframe is representative of the canonical 1-form on $G(M)$, we can then say that the effect of G on the canonical 1-form is a simple change of components in any horizontal tangent vector to $G(M)$ by a transformation in $G$:

$$(\Gamma \theta^i)(\mathbf{v}) = \Gamma (\theta^i(\mathbf{v})) = \Gamma_j^i v^j. \qquad (3.13)$$

The action of $\Gamma$ on G defined by:

$$(\Gamma, \omega) \infty \ \Gamma \omega \Gamma^{-1} + \Gamma d \Gamma^{-1}, \qquad (3.14)$$

which leads to an action of $\Gamma$ on the torsion and curvature 2-forms of $\omega$:

$$(\Gamma, \Theta) \infty \; \Gamma\Theta \tag{3.15a}$$

$$(\Gamma, \Omega) \infty \; \Gamma\Omega\,\Gamma^{-1}. \tag{3.15b}$$

Consequently, one can further decompose G into orbits of this action; i.e., gauge equivalence classes of g-connections. Such orbits are characterized by the fact that all of the connections in the same gauge orbit have conjugate curvature 2-forms. The isotropy subgroup $\Gamma_\omega$ of this action at any $\omega \in \Gamma$ is the subgroup of all $\Gamma$ that fix $\omega$, which leads to the condition:

$$\omega = \Gamma\omega\,\Gamma^{-1} + \Gamma\,d\,\Gamma^{-1}. \tag{3.16}$$

From this, we conclude that $\Gamma$ must be a constant function and its value must commute with $\omega$. Hence, the *G*-gauge orbits in G will generally look like $G/\Gamma_\omega$.

*c. Infinitesimal automorphisms.* One can also try to lift a vector field **v** on M to a vector field $\hat{\mathbf{v}}$ on $G(M)$ by an analogous process to the one that lifted a diffeomorphism. The difference is that when one picks an arbitrary $x \in M$ and looks at the local flow of **v** in a neighborhood $U$ of $x$:

$$\Phi_\tau : U \to M, \quad y \in U \infty \; \Phi_\tau(y), \tag{3.17}$$

with:

$$\left.\frac{d\Phi_\tau(y)}{d\tau}\right|_{\tau=0} = \mathbf{v}(y), \tag{3.18}$$

one sees that, in general, the linear isomorphism:

$$D\Phi_\tau\big|_x : T_x(M) \to T_{\Phi(x)}(M), \tag{3.19}$$

defines a corresponding diffeomorphism:

$$D\hat{\Phi}_\tau\big|_x : GL_x(M) \to GL_{\Phi(x)}(M), \tag{3.20}$$

but, unless $\Phi_\tau$ is also a *G*-isomorphism – at least at $\tau = 0$ – the lift of $\Phi_\tau$ to $GL(M)$ will not define a diffeomorphism:

$$D\hat{\Phi}_\tau\big|_x : G_x(M) \to G_{\Phi(x)}(M). \tag{3.21}$$

If this is indeed the case then one defines the lift of **v** to $G(M)$ at each $\mathbf{e}_x \in G(M)$ by:

$$\hat{\mathbf{v}}_{\mathbf{e}_x} = \left.\frac{d}{d\tau}\right|_{\tau=0} [D\hat{\Phi}_\tau\big|_x (\mathbf{e}_x)]. \tag{3.22}$$

This shows that only certain vector fields on *M* will lift to vector fields on *G(M)*. For instance, when $G = O(n)$ only Killing vector fields on *M* will lift to vector fields on *O(M)*. Note that if a lift to *G(M)* of a vector field on *M* exists then, by the linearity of $D\Phi_\tau\big|_x$, it will be $G$-invariant. We call a vector field on $M$ that lifts to a vector field on $G(M)$ an *infinitesimal automorphism* of $G(M)$.

Some equivalent criteria for a vector field $\mathbf{v}$ on $M$ to be an infinitesimal automorphism of a $G$-structure on $M$ are:

*a*)   If $\omega$ is a $G$-connection on $G(M)$ then $\omega_{\mathbf{e}}(\hat{\mathbf{v}}) \in \mathfrak{g}$ for all $\mathbf{e} \in G(M)$.

*b*)   If $\mathbf{e}_i$ is a local section of $G(U) \to U \subset M$ then:
$$L_{\mathbf{v}} \mathbf{e}_i = [\mathbf{v}, \mathbf{e}_i] = a\, \mathbf{e}_i, \quad \text{with} \quad a{:}U \to \mathfrak{g}.$$

*c*)   If $\theta^i$ is a local section of $G^*(U) \to U \subset M$ then:
$$L_{\mathbf{v}} \theta^i = b\, \theta^i, \quad \text{with } b{:}U \to \mathfrak{g}.$$

Note that when $\mathbf{e}_i$ and $\theta^i$ are reciprocal we must have $b = -a$.

*d*)   If $\theta^i$ is the canonical 1-form on $G(M)$ then:
$$L_{\hat{\mathbf{v}}} \theta^i = b\, \theta^i, \quad \text{with } b{:}G(M) \to \mathfrak{g}. \tag{3.23}$$

Suppose we have defined a $G$-connection on $G(M)$; as a consequence of *b*) and *c*), we have:

*e*)   If $\mathbf{z}$ is a horizontal vector field on $G(M)$ and $\hat{\mathbf{v}}$ is an infinitesimal automorphism then $[\hat{\mathbf{v}}, \mathbf{z}]$ is a vertical vector field.

Now suppose that the $G$-structure $G(M)$ is defined by a fundamental tensor field $t$. Since finite automorphisms always preserve $t$ it is not surprising that the necessary and sufficient condition for the lift of a vector field $\mathbf{v}$ on $M$ to a vector field $\hat{\mathbf{v}}$ on $GL(M)$ to be an infinitesimal automorphism of $G(M)$ is that:
$$L_{\hat{\mathbf{v}}}\, t = 0. \tag{3.24}$$

One could also formulate this as a necessary condition for a vector field $\hat{\mathbf{v}}$ on $G(M)$ to be the lift of some vector field on $M$; unless $\hat{\mathbf{v}}$ is also projectble, it is not however sufficient.

The set of all infinitesimal automorphisms of $G(M)$ defines a Lie subalgebra of $X(M)$; in particular, it is closed under Lie bracket. Furthermore, just as the group of vertical automorphisms of $G(M)$ was isomorphic to the group of smooth maps from $M$ to $G$, similarly, the Lie algebra of vertical infinitesimal automorphisms of $G(M)$ is isomorphic to the Lie algebra of smooth maps from $M$ to $\mathfrak{g}$. For the sake of consistency, we refer to the vertical infinitesimal automorphisms of a $G$-structure as its Lie algebra of *infinitesimal gauge transformations*.

An immediate special case of *d*) is when $b = 0$, i.e.:
$$L_{\hat{\mathbf{v}}} \theta^i = 0. \tag{3.25}$$

In such a case, we call the vector field $\hat{\mathbf{v}}$ an *infinitesimal parallelism.*

When one has defined a $G$-connection $\omega$ on $G(M)$, one can also define a vector field $\hat{\mathbf{v}}$ on $G(M)$ to be an *infinitesimal affine transformation* iff:

$$L_{\hat{v}}\,\omega = 0. \qquad (3.26)$$

*d. Infinitesimal gauge transformations and geometry.* When our infinitesimal automorphism projects to the identity, we can also look at how it alters the basic geometric data.

As usual, an infinitesimal $G$-gauge transformation must still preserve the fundamental tensor field:

$$L_{\tilde{a}}\,t = 0. \qquad (3.27)$$

However, from (3.23), if $\mathbf{a}: M \to \mathbf{g}$ and $\tilde{\mathbf{a}}\,(\mathbf{e}_x)$ is the fundamental vector at $\mathbf{e}_x$ that is defined by $\mathbf{a}(x)$ then the action of the Lie algebra $\Gamma(M, \mathbf{g})$ on is given by $\theta^i$:

$$L_{\tilde{a}}\,\theta^i = \mathbf{a}\,\theta^i, \qquad (3.28)$$

i.e., by infinitesimal $G$-transformations of the components of any horizontal vector, since:

$$(L_{\tilde{a}}\,\theta^i)(\mathbf{v}) = \mathbf{a}^i_{\ j}\,v^j. \qquad (3.29)$$

If we choose a smooth curve through the identity of $\Gamma(M, G)$ and differentiate the action (3.14) of $\Gamma$ on $\omega$ at the identity then we get the corresponding action of the infinitesimal gauge transformation on $\omega$:

$$L_{\tilde{a}}\,\omega = d\mathbf{a} + [\mathbf{a}, \omega], \qquad (3.30)$$

which leads to its action of torsion and curvature:

$$L_{\tilde{a}}\,\Theta^i = \mathbf{a}^i_{\ j}\,\Theta^j, \qquad (3.31a)$$

$$L_{\tilde{a}}\,\Omega = [\mathbf{a}, \Omega]. \qquad (3.31b)$$

Of course, since $Q = 0$, the effect of an infinitesimal gauge transformation on $Q$ is trivial.

In physical field theories, the primary role of infinitesimal gauge transformations is to generate field variations. If we understand that a variation of a physical field is basically the Lie derivative of that field with respect to the vector field that represents the infinitesimal generator of some fundamental deformation then we can rewrite the aforementioned equations as the definition of the variations in the geometrical objects that are produced by the variation $\tilde{\mathbf{a}}$ :

$$\delta t = 0, \qquad \delta\theta^i = \mathbf{a}^i_{\ j}\,\theta^j \qquad \delta\omega = d\mathbf{a} + [\mathbf{a}, \omega] \qquad (3.32a)$$

$$\delta Q = 0, \qquad \delta\Theta^i = \mathbf{a}^i_{\ j}\,\Theta^j \qquad \delta\Omega = [\mathbf{a}, \Omega]. \qquad (3.32b)$$

**4. Reducibility in the geometrical sequence.** Before we examine the specific nature of the various reductions in our geometrical subsequence, we first address the issue of whether the corresponding sequence of Lie algebras:

$$\mathfrak{a}(4) \leftarrow \mathfrak{gl}(4) \leftarrow \mathfrak{sl}(4) \leftarrow \mathfrak{so}(3,1) \leftarrow \mathfrak{so}(3) \leftarrow \mathbb{R} \leftarrow 0. \qquad (4.1)$$

is reducible at each step. As we shall see, each of the consecutive pairs of Lie algebras in this sequence, after the first one, is reductive. Although this fact can be deduced from more general statements about reducible Lie algebras, along the way we shall also describe the physical nature of the complementary subspaces and the decomposition itself.

$\mathbf{a}(4) \leftarrow \mathbf{gl}(4)$:    One knows that the affine group in $n$ dimensions is a semi-direct product of the general linear group in that dimension with the corresponding translation group. Hence, for Minkowski space, the affine Lie algebra decomposes as ([5]):

$$\mathbf{a}(4) = \mathbf{gl}(4) \& \mathbf{R}^4. \tag{4.2}$$

The Lie bracket obeys the laws:

$$\begin{aligned}
[\mathbf{gl}(4), \mathbf{gl}(4)] &\rightharpoonup \mathbf{gl}(4) \\
[\mathbf{gl}(4), \mathbf{R}^4] &\rightharpoonup \mathbf{gl}(4) \& \mathbf{R}^4 \\
[\mathbf{R}^4, \mathbf{R}^4] &= 0.
\end{aligned} \tag{4.3}$$

Although this does not exhibit $\mathbf{gl}(4)$ as reductive in $\mathbf{a}(4)$, that will not prove to be a geometrical obstacle, for other reasons.

$\mathbf{gl}(4) \leftarrow \mathbf{sl}(4)$:    The Lie algebra $\mathbf{sl}(4)$, which consists of 4×4 real matrices with trace zero, is reductive in $\mathbf{gl}(4)$, since any matrix in $\mathbf{gl}(4)$ commutes with any matrix of the form $\alpha I$; hence:

$$\mathbf{gl}(4) = \mathbf{R} \oplus \mathbf{sl}(4). \tag{4.4}$$

In particular, the decomposition of an element $a \in \mathbf{gl}(4)$ is:

$$a = \alpha I + a_0 \tag{4.5}$$

where:

$$\alpha = \tfrac{1}{4}\mathrm{Tr}(a), \quad a_0 = a - \tfrac{1}{4}\mathrm{Tr}(a)I; \tag{4.6}$$

hence, $a_0$ is the trace-free part of $a$. It represents the infinitesimal generator of a one-parameter subgroup of volume-preserving linear automorphisms of $\mathbf{R}^4$.

$\mathbf{sl}(4) \leftarrow \mathbf{so}(3,1)$: One has a decomposition of the Lie algebra $\mathbf{sl}(4) = \mathbf{s}_0 \oplus \mathbf{so}(3,1)$, where $\mathbf{s}_0$ is the vector space of *infinitesimal Lorentz strains*, or *Lorentz self-adjoint operators*, and $\mathbf{so}(3,1)$ is the Lie algebra of infinitesimal Lorentz transformations of Minkowski space. The decomposition is defined by Lorentz polarization:

$$a = \tfrac{1}{2}(a + a^*) + \tfrac{1}{2}(a - a^*), \tag{4.7}$$

where:

$$a^* = \eta a^{\mathrm{T}} \eta, \tag{4.8}$$

---

[5] Of course, the notation & refers to the semi-direct product of the Lie algebras.

and $\eta = \text{diag}[1, -1, -1, -1]$ is the matrix of the Lorentz scalar product on Minkowski space relative to the canonical basis for $\mathbf{R}^4$.

When one forms the Lie bracket $[l, \sigma]$ of an infinitesimal Lorentz transformation $l$ and an infinitesimal Lorentz strain $\sigma$, one obtains another infinitesimal Lorentz strain, since:

$$[l, \sigma]^* = (l\sigma - \sigma l)^* = \sigma^* l^* - l^* \sigma^* = -\sigma l + l\sigma = -[l, \sigma]; \qquad (4.9)$$

hence $\mathbf{so}(3,1)$ is reductive in $\mathbf{sl}(4)$.

$\mathbf{so}(3,1) \leftarrow \mathbf{so}(3)$: To examine the reducibility of $\mathbf{so}(3,1)$ to $\mathbf{so}(3)$, note that a common way of exhibiting the structure of the Lie algebra $\mathbf{so}(3,1)$ is by decomposing it as a vector space into a direct sum of infinitesimal boosts and infinitesimal spacelike rotations:

$$\mathbf{so}(3,1) = \mathbf{h} \oplus \mathbf{so}(3), \qquad (4.10)$$

with the commutations relations, which actually follow from the aforementioned polarization of elements of $GL(4)$:

$$\begin{aligned}
[\mathbf{so}(3), \mathbf{so}(3)] &\subset \mathbf{so}(3) \\
[\mathbf{so}(3), \mathbf{h}] &\subset \mathbf{h} \\
[\mathbf{h}, \mathbf{h}] &\subset \mathbf{so}(3,1).
\end{aligned} \qquad (4.11)$$

Hence, $\mathbf{so}(3)$ is reductive in $\mathbf{so}(3,1)$.

$\mathbf{so}(3) \leftarrow \mathbf{so}(2)$: If we choose a basis $\{I, J, K\}$ for $\mathbf{so}(3)$ whose first member $I$ generates the $\mathbf{so}(2)$-subalgebra in question then we can decompose any element $\omega \in \mathbf{so}(3)$ into a piece $\omega_1 I \in \mathbf{so}(2)$ another piece $\omega_2 J + \omega_3 K$. If we note that for any $aI \in \mathbf{so}(2)$:

$$[aI, bJ + cK] = ab[I, J] + ac[I, K] = abK - acJ \qquad (4.12)$$

then we see that if A is the vector subspace of $\mathbf{so}(3)$ that is spanned by $\{J, K\}$ then the decomposition $\mathbf{so}(3) = A \oplus \mathbf{so}(2)$ makes $\mathbf{so}(2)$ reductive in $\mathbf{so}(3)$. In effect, this is equivalent to expressing a three-dimensional rotation as a two-dimensional rotation about a particular axis; the space of such possible axes is parameterized by the 2-sphere.

## 5. The geometry of reductions in the geometrical sequence.

Although *G*-structures are generally reductions of the bundle $GL(M)$ of *linear* frames on the manifold *M*, we shall start with the bundle $A(M)$ of *affine* frames so we can account for the reduction of a connection on $A(M)$ to one on $GL(M)$ as a natural first step. It is not only very convenient that each reduction of our sequence involves a reductive pair of subalgebras, but, from Part I, we also see that each reduction is associated with a fundamental tensor field.

For each subgroup reduction of our geometrical subsequence, we shall examine the reduction of a connection from one subgroup to the next and the resulting geometry. In particular, we shall describe the data $\{t, \theta^\mu, \varpi\}$ and $\{Q, \Theta^\mu, \Omega\}$, the structure of the space of reduced connections, and the nature of the automorphisms of the structure ([6]).

$A(M) \to GL(M)$:    Since $A(4) = GL(4)\&\mathrm{R}^4$, the homogeneous space $A(4)/GL(4)$ is $\mathrm{R}^4$. In order to describe the reduction of $A(M)$ to $GL(M)$, we start with the fact that an *affine frame* in a tangent space $T_x(M)$ to $M$ consists of a pair $(\mathbf{p}, \mathbf{e}_\mu)$, where $\mathbf{p}$ is a point of $T_x(M)$ – or rather, its position vector relative to the origin – and $\mathbf{e}_\mu$ is a linear frame in $T_x(M)$. The fundamental tensor field $t$: $A(M) \to A(4)/GL(4) = \mathrm{R}^4$ for the reduction is defined by the map:

$$t: A(M) \to \mathrm{R}^4, \qquad (\mathbf{p}, \mathbf{e}_\mu)\in A_x(M) \infty \; p^\mu, \qquad (5.1)$$

where $\mathbf{p} = p^\mu \mathbf{e}_\mu \in T_x(M)$. Consequently, one can also regard the restriction of the fundamental tensor field $t$ to $GL(M)$ as equivalent to a vector field on $M$. A reduction from $A(M)$ to $GL(M)$ is then defined by the inverse image of a given vector in $\mathrm{R}^4$ under $t$, or a choice of vector field on $M$. For a given $x\in M$ and $a^\mu\in \mathrm{R}^4$, the frames of $t^{-1}(a^\mu)$ in $T_x(M)$ consist of all linear frames $\mathbf{e}_\mu$ in $T_x(M)$ that have their origin at $\mathbf{a} = a^\mu \mathbf{e}_\mu$.

It is clear how choosing a fixed vector field $\mathbf{p}: M \to T(M)$ defines a unique reduction from $A(M)$ to $GL(M)$ since a vector space is defined by an affine space and a point in that affine space that would serve as an origin; moreover, the vector space is the tangent vector space to the point in question. since the inverse image of $\mathbf{p}$ by $t$ in each fiber of $A(M)$ is the set of all linear frames $\mathbf{e}_\mu$ in $T_x(M)$ that are associated with an origin at $\mathbf{p}(x)$. Consequently, a reduction of $A(M)$ to $GL(M)$ is equivalent to a choice of vector field on $M$.

If $\theta^\mu$ is the linear coframe that is reciprocal to $\mathbf{e}_\mu$ then we can rewrite (5.1) as:

$$p^\mu: A(M) \to \mathrm{R}^4, \qquad (\mathbf{p}, \mathbf{e}_\mu)\in A_x(M) \infty \; \theta^\mu(\mathbf{p}). \qquad (5.2)$$

Now, if $\mathbf{p}$ were tangent to $GL(M)$ instead of $M$ then the map $p^\mu$ would be doing the same thing as the canonical 1-form $\theta^\mu$ on $GL(M)$. Hence, the restriction of $p^\mu$ to $GL(M)$ is equivalent to specifying the pair $(\mathbf{p}, \theta^\mu)$ where $\mathbf{p}$ is a vector field on $M$; for instance, one might simply use $\mathbf{p} = 0$.

Since any affine connection $\omega_\nu^\mu$ on $A(M)$ takes its values in $a(4) = gl(4)\&\mathrm{R}^4$, it can be decomposed into the sum $\omega = \varpi + \theta$ of a 1-form $\varpi$ with values in $gl(4)$ and a 1-form $\theta$ with values in $\mathrm{R}^4$. If we use $(0, \theta)$ to define $p^\mu$ then $\varpi$ will be the 1-form for a linear connection $\varpi_\nu^\mu$ and $\theta$ will be the canonical 1-form $\theta^\mu$ on $GL(M)$, respectively.

If we represent the Lie algebra of $A(4)$ by $5\underline{\times}5$ real matrices in semi-direct product form, we can say that:

---

$$\omega_\nu^\mu = \left[\begin{array}{c|c} \tilde{\omega}_\nu^\mu & \theta^\mu \\ \hline 0 & 1 \end{array}\right]. \tag{5.3}$$

Note that because $\omega$ is a linear isomorphism of vertical tangent spaces on $A(M)$ with $a(4)$, one has that $V(A(M))$ is trivializable ([7]). Since the pair $(\theta, \varpi)$ collectively defines a linear isomorphism of each tangent space to $GL(M)$ with $a(4)$, as well, one concludes that $T(GL(M))$ is isomorphic to $V(A(M))$, which makes $T(GL(M))$ trivializable by way of $(\theta, \varpi)$. Hence, if one defines the one-form $\omega = \varpi + \theta$ with values in $a(4)$ on $GL(M)$, instead of $A(M)$, then $\omega$ defines a linear isomorphism of each tangent space $T_e GL(M)$ with the vector space $gl(4) \oplus \mathbb{R}^4$, which also defines a parallelization of $GL(M)$. One then refers to $\omega$ as a (special type of) *Cartan connection* [**9**] on $GL(M)$.

A glance at (5.3) shows that the only way that an affine connection is *directly* reducible to a linear connection is if the canonical 1-form $\theta^\mu$ on $GL(M)$ is identically zero, which is absurd, but one can see that any choice of value for **p** *defines* a linear connection on $GL(M)$ that corresponds to any affine connection on $A(M)$. In fact, for a given choice of $p^\mu$ − i.e., a given choice of **p** − the correspondence between $\omega$ and $\varpi$ is one-to-one.

The irreducibility 1-form $Q$ becomes the 1-form on $A(M)$ with values in $\mathbb{R}^4$:

$$Q^\mu = \nabla p^\mu, \tag{5.4}$$

which vanishes iff the vector field **p** on $M$ is parallel with respect to $\omega$.

The curvature 2-form of $\omega$ splits in a manner that is analogous to (5.3):

$$\Omega_\nu^\mu = \left[\begin{array}{c|c} \tilde{\Omega}_\nu^\mu & \Theta^\nu \\ \hline 0 & 1 \end{array}\right]. \tag{5.5}$$

The affine space G of $gl(4)$-connections on $GL(M)$ is distinguished from the orthogonal subgroups of the geometrical sequence by having a subspace of torsionless connections that consists of more than one element. As pointed out above, the vector space on which it is modeled is the space $G_0$ of $GL(4)$-equivariant $gl(4)$-valued 1-forms $\tau$ on $GL(M)$ such that $\tau \wedge \theta = 0$. As a module over the ring $C^\infty(GL(M))$ this space is 40-dimensional. In effect, $G_0$ is the subspace of torsion-preserving affine transformations of the linear connections.

An automorphism of $GL(M)$ then becomes a $GL(4)$-invariant diffeomorphism $f$ of $A(M)$ such that $f^* p^\mu$, which means that it translates linear frames along the flow of the vector field **p** in such a way that they always have their origin at the corresponding point defined by **p** at the translated point of $M$. An infinitesimal generator for such an automorphism is then a *G*-invariant vector field $\hat{\mathbf{v}}$ on $A(M)$ such that $\hat{\mathbf{v}}p^{\mu} = 0$, which says that $p^{\mu}$ is constant in the direction $\hat{\mathbf{v}}$.

---

[7] Of course, one has, more generally, that since the vertical sub-bundle of $T(P)$ for a $G$-principal bundle $P \rightarrow M$ can always be framed by the fundamental vector fields defined any basis for g the vertical sub-bundle to a $G$-principal bundle is always trivializable.

The gauge transformations of $GL(M)$ are smooth maps from $M$ to $GL(4)$ and their infinitesimal generators are smooth maps from $M$ to $\mathfrak{gl}(4)$.

$GL^{+}(M) \to SL(M)$:        The homogeneous space $GL^{+}(4)/SL(4)$ is diffeomorphic to $\mathbf{R}^{+}$, since all we are doing is factoring any invertible $4 \times 4$ real matrix with positive determinant $A$ into $\det(A)[\det(A)^{-1}A]$.

The fundamental tensor field *t* for this reduction is the $GL^{+}$ (4)-equivariant map:

$$\det: GL^{+}(M) \to \mathbf{R}^{+}, \mathbf{e} \propto \det(\mathbf{e}), \qquad (5.6)$$

which is essentially am $SL(4)$-invariant real-valued 0-form on $GL^{+}(M)$. A reduction from $GL+(M)$ to a choice of $SL(M)$ is then defined by a choice of $a \in \mathbf{R}^{+}$ and the set $\det^{-1}(a)$.

Typically, one might use $a = 1$, but one should keep in mind that since all of the other reductions that are defined by other choices of *a* are isomorphic there is nothing special about the number 1 in $\mathbf{R}^{+}$, any more than most affine spaces have a privileged point to define an origin. This freedom to choose a "unit volume" at each point arbitrarily was originally suggested by Weyl and Eddington [**10, 11**] as the source of the $U(1)$-gauge invariance of electromagnetism, although one sees that since $U(1) = SO(2)$ is the one-point compactification of $\mathbf{R}^{+}$, the isomorphism applies only at the infinitesimal level.

The irreducibility tensor for a connection $\omega$ on $GL^{+}(M)$ is then ([8]):

$$Q = \nabla t = \nabla(\det(\mathbf{e}_{\mu})) = D(\det)(\nabla \mathbf{e}_{\mu}) = -D(\det)(\omega_{\mu}^{\nu} \otimes \mathbf{e}_{\nu})$$
$$= d(\ln \det(\mathbf{e}_{\mu})) - \mathrm{Tr}(\omega_{\nu}^{\mu}), \qquad (5.7)$$

whose restriction to $SL(M)$ is:

$$\mathrm{Tr}(\omega) = -\mathrm{Tr}(\omega_{\nu}^{\mu}). \qquad (5.8)$$

If $\omega$ reduces to an $\mathfrak{sl}(4)$-connection on $SL(M)$ then this must vanish.

Since $\mathfrak{sl}(4)$ is reductive in $\mathfrak{gl}(4)$, any connection $\omega$ on $GL^{+}(M)$ can be decomposed into an $\mathfrak{sl}(4)$-connection $\varpi$ on $SL(M)$ and a deformation 1-form $\tau_1$ whose values take the form of an ordinary 1-form times the $4 \times 4$ identity matrix

$$\omega = \varpi + \tau_1, \qquad (5.7)$$

where:

$$\varpi = \omega - \tfrac{1}{4}\mathrm{Tr}(\omega)I = \omega + \tfrac{1}{4}QI, \qquad \tau_1 = \tfrac{1}{4}\mathrm{Tr}(\omega)I = -\tfrac{1}{4}QI. \qquad (5.8)$$

Note that, from Schur's lemma, the $\mathbf{R}$-factor in the decomposition $\mathfrak{gl}(4) = \mathfrak{sl}(4) \oplus \mathbf{R}$ is the center of $\mathfrak{gl}(4)$.

---

[8] We use the notation *D* for the differential map to distinguish it from the exterior derivative, which we denote by *d*.

The affine space of $\mathfrak{sl}(4)$-connections is modeled on the vector space $\Lambda^1_{eq}(SL(M); \mathfrak{sl}(4))$, which differs from the space of $\mathfrak{gl}(4)$-connections only by a factor of $\Lambda^1_{eq}(SL(M); \mathbb{R}) = \Lambda^1(M)$, since the decomposition (4.7) give us that:

$$\Lambda^1_{eq}(SL(M); \mathfrak{gl}(4)) = \Lambda^1_{eq}(SL(M); \mathfrak{sl}(4)) \oplus \Lambda^1_{eq}(SL(M); \mathbb{R}). \qquad (5.9)$$

The automorphisms of $SL(M)$ are going to be the lifts of volume-preserving diffeomorphisms of $M$. Their infinitesimal generators are then going to be $SL(4)$-invariant vector fields $\hat{\mathbf{v}}$ on $SL(M)$ such that:

$$L_{\hat{\mathbf{v}}} \det = i_{\hat{\mathbf{v}}} d(\det) = \text{Tr}(\hat{\mathbf{v}}) = 0, \qquad (5.10)$$

in which the map Tr is defined as the composition Tr: $T(GL^+(M)) \to \mathfrak{gl}(4) \to \mathbb{R}$, $\hat{\mathbf{v}} \infty a^\mu_\nu(\hat{\mathbf{v}}) \infty a^\mu_\mu(\hat{\mathbf{v}})$.

If we represent the fundamental tensor field for the reduction as a volume element on the vector bundle $H(G(M))$:

$$\mathbf{\vartheta} = \det(\mathbf{e})\theta^0{\wedge}\,\theta^1{\wedge}\,\theta^2{\wedge}\,\theta^3 = \frac{1}{4!}\det(\mathbf{e})\,\varepsilon_{ijkl}\,\theta^i{\wedge}\,\theta^j{\wedge}\,\theta^k{\wedge}\,\theta^l. \qquad (5.11)$$

then we can also characterize the infinitesimal generators of $SL(M)$ automorphisms as the lifts of divergenceless vector fields on $M$.

$SL(M) \leftarrow SO(3, 1)(M)$:   The homogeneous space $\Sigma(4) = SL(4)/SO(3,1)$ consists of volume-preserving Lorentz shears. These are $4{\cdot}4$ real matrices $\Sigma$ with unit determinant that are self-adjoint with respect to the Lorentz scalar product:

$$\Sigma^* = \Sigma, \qquad\qquad \Sigma^* = \eta\Sigma^T\eta; \qquad (5.12)$$

in particular, $\eta_{\mu\nu} \in \Sigma(4)$. The tangent vectors to $\Sigma(4)$ then consist of self-adjoint matrices with zero trace.

The fundamental tensor field of this reduction is the metric tensor field $g_{\mu\nu}$. For a Lorentzian frame $g_{\mu\nu}(\mathbf{e}) = \eta_{\mu\nu}$. However, just as any oriented linear frame could be defined to have unit volume, similarly, any unit volume frame could be defined to be Lorentzian. In either case the choice of reduction amounts to the choice of orbit through the frame in question.

The irreducibility tensor for a connection $\omega$ on $SL(M)$ is then an $SL(4)$-equivariant 1-form on $SL(M)$ with values in $\Sigma(4)$ that is commonly called the *nonmetricity* of $\omega$:

$$Q_{\mu\nu} = \nabla g_{\mu\nu} = dg_{\mu\nu} - \omega^\lambda_\mu\, g_{\lambda\nu} - \omega^\lambda_\nu\, g_{\lambda\mu}. \qquad (5.13)$$

The condition for the restriction of an $\mathfrak{sl}(4)$-connection $\omega$ to $SO(1,3)(M)$ to reduce to an $\mathfrak{so}(3,1)$-connection is the familiar constraint that $\omega$ must be a metric connection:

$$Q = 0. \qquad (5.14)$$

For Lorentzian frames this implies that:

$$\omega_{\mu\nu} + \omega_{\nu\mu} = 0. \tag{5.15}$$

One can solve the equation $\tau \wedge t = Q$ for $t$ by way of symmetrization:

$$\tau^{\lambda}_{\mu\nu} = \tfrac{1}{2} g^{\lambda\sigma} (Q_{\sigma;\mu\nu} + Q_{\mu;\lambda\nu} + Q_{\nu;\lambda\mu}) \tag{5.16}$$

in manner that is completely analogous to the way that one solves for contortion 1-form in terms of the torsion 2-form.

Of course, the space of torsionless metric connections consists of one point, namely the Levi-Civita connection. Hence, the vector space on which the affine space of metric connections is simply $\Lambda^{1}_{eq}(SL(M); \mathfrak{so}(3,1))$.

The automorphisms of $SO(1,3)(M)$ are the lifts of Lorentzian isometries on $M$, which are also affine transformations. Their infinitesimal generators are the lifts of Killing vector fields on $M$.

$SO_0(3,1)(M) \to SO(3)(M)$:    The homogeneous space $SO_0(3,1)/SO(3) = \mathbf{R}^3$ represents the space of possible directions for infinitesimal boosts. The fundamental tensor field of this reduction is an $SO_0(3,1)$-equivariant map $t^{\mu}: SO_0(3,1)(M) \to \mathbf{R}^3$ that gives rise to the timelike unit vector field $\mathbf{t} = t^{\mu}\mathbf{e}_{\mu}$ on $M$, which represents the direction of proper time evolution at each point. We point out that in this Part of our study, we are combining the reduction from oriented Lorentzian frames to oriented Lorentzian frames that have one member that generates the same *line* as $\mathbf{t}$, which represents a *pair* of Lorentzian frames at each point, with the reduction of the pair to a unique frame. In other words, we have trivialized the line bundle $L(M)$, whose existence is equivalent to defining a Lorentzian metric, by defining an orientation for it, which is generally referred to as a *time orientation*. Of course, as we discussed in Part I this is potentially obstructed by the topology of $M$ and the nature of the bundle $L(M)$.

Now that we have a Lorentzian metric, we can speak of orthogonal decompositions in the tangent spaces. We have a Whitney sum decomposition:

$$T(M) = \mathbf{R}(M) \oplus \Sigma(M), \tag{5.17}$$

in which $\mathbf{R}(M)$ is the trivial timelike line bundle spanned by $\mathbf{t}$ and $\Sigma(M)$ is its spacelike complement. Similarly, one can decompose the Lie algebra of vector fields on $M$ into *vector subspaces*:

$$\mathbf{X}(M) = \mathbf{X}_t(M) \oplus \mathbf{X}_{\Sigma}(M), \tag{5.18}$$

so any vector field can be decomposed into a timelike and a spacelike part. Moreover, since $\mathbf{R}(M)$ is trivial, any vector field of $\mathbf{X}_t(M)$ can be uniquely represented by a smooth map from $M$ to $\mathbf{R}$. One is warned that, by Frobenius, the decomposition (5.1) does not have to represent a decomposition into Lie subalgebras, unless both sub-bundles are integrable. Since the fibers of $\Sigma(M)$ are one-dimensional, this is straightforward, but for $\mathbf{X}_{\Sigma}(M)$ it is not. (For more discussion of the integrability of $\Sigma(M)$, see [12].) Note that although the fibers of $\mathbf{R}(M)$ are one-dimensional, nevertheless the Lie algebra $\mathbf{X}_t(M)$ is neither finite-dimensional nor Abelian.  Indeed, for arbitrary sections of R(M) of the form α**t**, β**t**, one has:

$$[\alpha\mathbf{t}, \beta\mathbf{t}] = \{\alpha(\mathbf{t}\beta) - \beta(\mathbf{t}\alpha)\}\mathbf{t}. \tag{5.19}$$

In fact, this allows us to define the same Lie algebra on the vector space of smooth functions on M by way of:

$$[\alpha\mathbf{t}, \beta\mathbf{t}] = \alpha(\mathbf{t}\beta) - \beta(\mathbf{t}\alpha), \tag{5.20}$$

which is reminiscent of the construction of the Poisson algebra for the smooth functions on a symplectic manifold.

Since the sub-bundle Σ(M) defines a rank-three differential system, depending upon its integrability one could also say that the reduction to a rest frame is also a reduction in *dimension* from spacetime to a *rest space*, which we understand to be a three-dimensional (proper-time simultaneity) leaf of the foliation ([9]).  One observes that rest spaces seem to play a role in relativistic dynamics that is closely analogous to the role of static solutions to time-varying differential equations.  One can also regard the geometry of an SO(3)-structure on spacetime as being the extrinsic geometry of the leaves when regarded as submanifolds.

One sees that the three-dimensional Euclidean space on which SO(3) acts at each point of spacetime is Σ(M).  The frames of SO(3)(M) are then all of the oriented Lorentzian 4-frames that share a common member, namely **t**.  The orthogonal frames of SO(3)(M) are then referred to as *rest frames* for the motion defined by **t**.  As a result, the matrices that act on the Lorentzian frames are reduced to the SO(3) subgroup of Lorentzian matrices of the form:

$$\begin{bmatrix} 1 & 0 \\ 0 & R_3 \end{bmatrix}, \tag{5.21}$$

in which $R_3$ is a 3ꓷ3 orthogonal matrix.  Because of this, one generally omits the timelike part and simply deals with the action of $R_3$ on the spacelike 3-frames in Σ(M).  Similarly, the Lie algebra so(3) either gets represented by 4ꓷ4 matrices of the form:

$$\begin{bmatrix} 0 & 0 \\ 0 & \omega_3 \end{bmatrix}, \tag{5.22}$$

or, more concisely, by the 3ꓷ3 antisymmetric part $\omega_3$.  The consequences of (5.21) and (5.22) in terms of finite and infinitesimal gauge transformations are immediate.

The irreducibility tensor for a metric connection ω on $SO_0(3,1)(M)$ is then:

$$Q^\mu = \nabla t^\mu \tag{5.23}$$

When this tensor vanishes one sees that **t** must be a timelike vector field that is parallel for the connection ω.  One can either regard $Q^\mu$ as being the covariant acceleration of the motion that defines **t**, or if the differential system $\Sigma(M)$ is integrable then one can also regard $Q^\mu$ as being the geodesic curvature of the leaves.

The decomposition of a Lorentzian connection $\omega$ into the sum of an $\mathfrak{so}(3)$-connection $\varpi$ and a difference form $\tau$ with values in the vector space b of infinitesimal boosts is by Lorentz polarization:

$$\varpi = \tfrac{1}{2}(\omega - \omega^*), \quad \tau = \tfrac{1}{2}(\omega + \omega^*), \tag{5.24}$$

where the Lorentzian adjoint of $\omega$ is:

$$\omega^* = \eta \omega^\mathrm{T} \eta. \tag{5.25}$$

Hence, we can also deduce the compatibility condition for the reduction from (5.23) and (5.24):

$$\nabla t^\mu = \tfrac{1}{2}(\omega + \omega^*)^\mu_\nu \wedge t^\nu. \tag{5.26}$$

The space of $\mathfrak{so}(3,1)$-connections on $SO(3,1)(M)$ decomposes accordingly into the space of $\mathfrak{so}(3)$-connections on $SO(3,1)(M)$ and the space of $SO(3)$-invariant 1-forms on $SO(3,1)(M)$ with values in b. The space of torsionless $\mathfrak{so}(3)$-connections is a single point, namely the restriction to $SO(3)(M)$ of the Levi-Civita connection for the pseudo-metric on $T(M) = \mathrm{R}(M) \oplus \Sigma(M)$ that is defined by:

$$g' = \theta \otimes \theta - g_\Sigma, \tag{5.27}$$

in which $\theta$ is the metric-dual 1-form to **t** and $g_\Sigma$ is the restriction of our Lorentzian $g$ to the spacelike sub-bundle $\Sigma(M)$.

The automorphisms of $SO(3)(M)$ are going to define the subgroup of the automorphisms of $SO(3,1)(M)$ that fix the **t** member of the reduced Lorentzian frames. In effect, these transformations are composed of isometries of the rest spaces and parallel convections along the flow of **t**. Their infinitesimal generators will be the lifts of Killing vector fields on $M$ that are orthogonal to **t** plus vector fields of the form $\alpha$**t**, where $a$ is a smooth function that constant in the direction **t**; in the integrable case, the infinitesimal isometries of the rest spaces will be tangent to the leaves.

$SO(3)(M) \to SO(2)(M)$: The homogeneous space for this reduction is the 2-sphere $SO(3)/SO(2)$, which represents the set of all unit vectors in $\mathrm{R}^3$, which we understand to be the model space for $\Sigma(M)$.

The fundamental tensor field for this reduction:

$$n^i: SO(3)(M) \to S^2, \quad \mathbf{e}_\mu \leadsto n^i(\mathbf{e}_\mu) \tag{5.28}$$

is equivalent to the spacelike unit vector field $\mathbf{n} = n^i \mathbf{e}_i$, which is a section of the vector bundle $\Sigma(M)$. Physically, we interpret $\mathbf{n}$ as the unit normal vector field to a codimension-one foliation of $M$ by timelike isophase hypersurfaces (cf. [**2**]) for a wave motion, depending upon the integrability of the differential system defined by the rank-3 sub-bundle $\Phi(M)$ of $T(M)$ defined by the orthogonal complement to the line bundle spanned by **n**.

One can also define the intersection $W(M) = \Sigma(M) \cap \Phi(M)$ of the two codimension-one sub-bundles $\Sigma(M)$ and $\Phi(M)$. If $W$ is integrable then its leaves are two-dimensional submanifolds called *momentary wavefronts*, which are intersections of isophase leaves with simultaneity leaves, and whose tangent spaces are the intersections of the corresponding rank-3 subspaces in $\Sigma(M)$ and $\Phi(M)$. Hence, the geometry of $SO(2)(M)$ can also be regarded as relating to the geometry of momentary wavefronts.

This latter constructions of the rank-2 spacelike sub-bundle $W(M)$, together with the trivial rank-2 timelike sub-bundle $R^2(M)$ of $T(M)$ that is spanned by the frame field $\{\mathbf{t}, \mathbf{n}\}$ defines a Whitney sum splitting:

$$T(M) = R^2(M) \oplus W(M), \tag{5.29}$$

and with it a splitting of the Lie algebra of vector fields into vector subspaces:

$$X(M) = X_m(M) \oplus X_W(M). \tag{5.30}$$

Hence, any vector field on $M$ can be decomposed into a section of $W(M)$ and a section of $R^2(M)$. Since $R^2(M)$ is trivial one can also represent its elements as smooth functions from $M$ to $R^2$. Once again, whether the vector spaces $X_m(M)$ and $X_W(M)$ define Lie subalgebras of $X(M)$ is equivalent to the integrability of the differential systems $R^2(M)$ and $W(M)$. In the integrable case, the vector fields of $X_W(M)$ will be tangent to the momentary wavefronts.

The splitting (5.29) also implies that the matrices that act on the frames will take the form:

$$\begin{bmatrix} 1 & 0 & 0 \\ 0 & 1 & 0 \\ 0 & 0 & R_2 \end{bmatrix}, \tag{5.31}$$

in which $R_2$ is a $2 \times 2$ rotation matrix, or, more concisely, by $R_2$ and a suitable restatement of the action of such matrices on 4-frames, as we did for the previous reduction. Similarly, the Lie algebra $so(2)$ can either be represented by matrices of the form:

$$\begin{bmatrix} 0 & 0 & 0 \\ 0 & 0 & 0 \\ 0 & 0 & \omega_2 \end{bmatrix}, \tag{5.32}$$

or by the $2 \times 2$ antisymmetric matrix $\omega_2$ and a suitable definition of its action on tangent vectors to the frames. Clearly, the $2 \times 2$ matrices $R_2$ and $\omega_2$ will also define the most efficient representations for the gauge transformations of $SO(2)(M)$ and their infinitesimal generators, respectively.

The irreducibility tensor for the restriction of an $so(3)$-connection $\omega$ on $SO(3)(M)$ to $SO(2)(M)$ is then the spacelike one-form with values in $R^3$:

$$Q^i = \nabla n^i = dn^i + \omega^i_j \wedge n^j, \tag{5.33}$$

which is equivalent to the second-rank tensor field $\nabla \mathbf{n} = \nabla n^i \otimes \mathbf{e}_i$ on $M$.

The condition for a time-oriented spacelike metric connection $\omega$ on $SO(3)(M)$ to reduce to an $so(2)$-connection on $SO(2)(M)$ is then that $\nabla \mathbf{n} = 0$, i.e., that $\mathbf{n}$ is parallel under the $so(3)$-connection on $SO(3)(M)$. Hence, the orthonormal 2-frame field $\{\mathbf{t}, \mathbf{n}\}$ must be parallel with respect to the $so(3,1)$-connection on $SL(M)$. Since $\mathbf{n}$ is a unit vector one can solve the equation:

$$\tau^i_j n^j = Q^i \tag{5.34}$$

for $\tau^i_j$ by way of:

$$\tau^i_j = g_{jk} n^k Q^i = n_j Q^i. \tag{5.35}$$

This solution is not unique, since one could add a term of the form $b_j Q^i$ to it, where $\mathbf{b}$ is orthogonal to $\mathbf{n}$.

In the integrable case, $Q^i$ represents the geodesic curvature of the momentary wavefronts as submanifolds of the proper-time simultaneity leaves or the covariant acceleration of the flow defined by $\mathbf{n}$. However, since $\mathbf{n}$ does not play the physical role of a generator for a physical, one must realize the flow of $\mathbf{n}$ has more to do with the structure of the geodesics of the momentary wave fronts. In particular, one often prefers that these submanifolds be *totally geodesic*; i.e., any geodesic that starts off in the submanifold remains in the submanifold.

Because $SO(2)$ is one-dimensional an $so(2)$-connection $\varpi$ on $SO(2)(M)$ can be factored into a product:

$$\varpi^i_j = \varpi J^i_j, \tag{5.36}$$

where:

$$J^i_j = \begin{bmatrix} 0 & -1 \\ 1 & 0 \end{bmatrix}. \tag{5.37}$$

Similarly, one can simplify the expressions for the torsion and curvature of $\varpi$ into 2-forms, $\Theta^i$ and $\Omega$. We write the equations for the individual components explicitly for the sake of clarity:

$$\Theta^1 = d\theta^1 - \varpi \wedge \theta^2, \qquad \Theta^2 = d\theta^2 + \varpi \wedge \theta^1, \qquad \Omega = d\varpi = \kappa \, \theta^1 \wedge \theta^2. \tag{5.38}$$

When the 2-form $\Theta^i$ vanishes the first two equations give the *Gauss-Codazzi* equations for the momentary wavefronts, and the 0-form $\kappa$ represents its *Gaussian curvature* (see O'Neill [13] for surface geometry described in these terms).

The space of $so(2)$-connections on $SO(2)(M)$ is modeled on the vector space $\Lambda^1_{eq}(SO(2)(M); R) = \Lambda^1(M)$. It, too, has a unique torsionless element in the form of the Levi-Civita connection $[\varpi_0]^i_j$, which simplifies into a product:

$$[\varpi_0]^i_j = \varpi_0 J^i_j, \tag{5.39}$$

and $\varpi_0 = \alpha\theta^1 + \beta\theta^2$ can be obtained from the Gauss-Codazzi equations, which become:

$$d\theta^1 = \alpha \ \theta^1 {}^\wedge \ \theta^2, \qquad d\theta^2 = \beta \ \theta^1 {}^\wedge \ \theta^2. \qquad (5.40)$$

One can use the matrix $J_j^i$ to define an almost-complex structure $J: M \to W^*(M) \otimes W(M)$ on the bundle $W(M)$ by using its action on the frames of $SO(2)(M)$. If $\mathbf{v} \in W_x(M)$ and $\mathbf{e}_i \in SO(2)_x(M)$, and we confine our attention to the members of the frame that are in $W_x(M)$ then this action is defined by:

$$J(\mathbf{v}) = v^j J_j^i \mathbf{e}_i. \qquad (5.41)$$

The fact that this definition is independent of the choice of frame is due to the fact that $SO(2)$ is Abelian, so $J$ commutes with all of the allowable rotations of $\mathbf{e}_i$.

The curvature 2-form $\Omega$ can also be regarded as proportional to the first Chern class of the bundle $W(M)$. Hence, whether this bundle is indeed trivial will depend upon its vanishing. Moreover, by Gauss-Bonnet, the integral of $\Omega$ over any momentary wavefront $S$ must not only be proportional to the Euler-Poincaré characteristic $\chi[S] = 2 - 2g$, where $g$ is the genus of $S$, but this number can take only a denumerable sequence of values.

The automorphisms of $SO(2)(M)$ will be the lifts of isometries on $M$ that fix $\mathbf{t}$ and $\mathbf{n}$, either transversally or along their flows. In the integrable case, these will be isometries of the momentary wave fronts and other parallel translations of the frame $\{\mathbf{t}, \mathbf{n}\}$. The infinitesimal generators will then the lifts of Killing vector fields that are orthogonal to $\mathbf{t}$ and $\mathbf{n}$ and vector fields of the form $\mathbf{v} = \alpha\mathbf{t} + \beta\mathbf{n}$. However, this time, since $\mathbf{v}$ must commute with *both* $\mathbf{t}$ and $\mathbf{n}$, and:

$$[\mathbf{v}, \mathbf{t}] = -(\mathbf{t}\alpha)\mathbf{t} - \beta[\mathbf{t}, \mathbf{n}] - (\mathbf{t}\beta)\mathbf{n} \qquad (5.42a)$$

$$[\mathbf{v}, \mathbf{n}] = -(\mathbf{n}\alpha)\mathbf{t} + \alpha[\mathbf{t}, \mathbf{n}] - (\mathbf{n}\beta)\mathbf{n} \qquad (5.42a)$$

the form that $\mathbf{v}$ must take depends upon whether $[\mathbf{t}, \mathbf{n}] = 0$ or not. If $\mathbf{t}$ and $\mathbf{n}$ commute then $\alpha$ and $\beta$ can be any smooth functions that vary only in the directions orthogonal to both $\mathbf{t}$ and $\mathbf{n}$. This possibility is equivalent to the 2-frame field $\{\mathbf{t}, \mathbf{n}\}$ being locally integrable into the natural frame field for a coordinate chart on a 2-dimensional integral submanifold of the differential system defined by $\mathbf{R}^2(M)$. If $[\mathbf{t}, \mathbf{n}]$ do not commute then one must further subdivide the possibilities depending upon whether this bracket is still a vector field in $X_m(M)$ or not. If so, then we are still dealing with the case of an integrable $\mathbf{R}^2(M)$, but this time the 2-frame $\{\mathbf{t}, \mathbf{n}\}$ is not natural. If we let $[\mathbf{t}, \mathbf{n}] = \rho\mathbf{t} + \sigma\mathbf{n}$ then (5.42a,b) imply that we must have:

$$0 = -(\mathbf{t}\alpha + \beta\rho)\mathbf{t} - (\mathbf{t}\beta + \beta\sigma)\mathbf{n} \qquad (5.43a)$$

$$0 = (\alpha\sigma - \mathbf{n}\alpha)\mathbf{n} + (\alpha\rho - \mathbf{n}\beta)\mathbf{t} \qquad (5.43a)$$

which gives the system of partial differential equations for $\alpha$ and $\beta$:

$$\mathbf{t}\alpha + \beta\rho = 0, \qquad\qquad \mathbf{t}\beta + \beta\sigma = 0 \qquad (5.44a)$$

$$\mathbf{n}\alpha - \alpha\sigma = 0, \qquad\qquad \mathbf{n}\beta + \alpha\rho = 0. \qquad (5.44a)$$

In the more general case where [**t**, **n**] also has a contribution **w** from the vectorfields in $X_W(M)$, one must also consider the way that **w** commutes with both **t** and **n**, and the resulting system of equations expands accordingly. Such systems of partial differential equations whose solution give infinitesimal automorphisms of *G*-structures are called, more generally, *Lie equations*, although a discussion of such matters is far beyond the intended scope of the immediate study.

$SO(2)(M) \rightarrow \{e\}(M)$:  The homogeneous space for this reduction is the unit circle, which represents all unit vectors in $R^2$, which we understand to be the model space for the spacelike rank-two vector bundle $W(M)$.

Consequently, the fundamental tensor field of this reduction:

$$p^i: SO(2)(M) \rightarrow S^1 \hookrightarrow R^2, \qquad \mathbf{e}_\mu \propto p^i(\mathbf{e}_\mu) \qquad (5.45)$$

is equivalent to the spacelike unit vector field $\mathbf{p} = p^i \mathbf{e}_i$, $i = 1, 2$. Physically, the line spanned by **p** at each point $x \in M$ represents a choice of "zero phase" line in $W_x(M)$ for the action of $SO(2)$; once again, we are overlooking the orientability issues that we discussed in Part I. One could say that we have defined a further splitting of $T(M)$ into the Whitney sum:

$$T(M) = R^3(M) \oplus L(M), \qquad (5.46)$$

where $R^3(M)$ is the trivial bundle that is spanned by the orthonormal frame 3-field {**t**, **n**, **p**} and $L(M)$ is a spacelike line bundle whose lines belong to $W(M)$, or rather, its projectivization. However, since we have a Lorentzian metric and an orientation on $T(M)$, we can complete the orthonormal triad {**t**, **n**, **p**} to an orthonormal tetrad {**t**, **n**, **p**, **q**}, i.e., a global frame field on $M$, by choosing $L(M)$ to be orthogonal to $R^3(M)$ and the unit vector field **q** in $L(M)$ that makes the resulting tetrad consistent with the choice of orientation. Hence, we have trivialized $T(M)$ completely into $R^4(M)$, so the vector fields can be uniquely associated with smooth maps from $M$ to $R^4$.

The irreducibility tensor field for an $\mathfrak{so}(2)$-connection $\omega$ on $SO(2)(M)$ is then the 1-form with values in $R^2$:

$$\nabla p^i = dp^i + \omega^i_j p^j = dp^i + \omega \ J^i_j p^j \qquad (5.47)$$

which is equivalent to the second rank tensor field on $M$ defined by $\nabla \mathbf{p} = \nabla p^i \otimes \mathbf{e}_i$, $i = 1, 2$.

The condition for an $\mathfrak{so}(2)$-connection $\omega$ on $SO(2)(M)$ to reduce to an $\{e\}$-connection $\varpi$ on $\{e\}(M)$ is then that the vector field **p** must be parallel with respect to $\omega$. However, we immediately point out the Lie algebra in which the $\{e\}$-connection $\varpi$ takes its values is simply $\{0\}$ and an $\{e\}$-structure is a choice of global frame field. Hence, for any frame in $\{e\}(M)$ the reducibility condition reduces to:

$$\nabla p^i = dp^i = 0. \qquad (5.49)$$

which says that the components $p^i$ of **p** with respect to that frame are constant. Of course, this is how one defines parallelism by means of a global frame field.

The 1-form $\tau$ must then satisfy:

$$\tau \, J^i_j \, p^j = dp^i, \tag{5.50}$$

or:

$$\tau \, p^2 = -dp^1, \tag{5.51a}$$

$$\tau \, p^1 = +dp^2, \tag{5.51b}$$

which can always be solved for $\tau$ since $\mathbf{p}$ is a unit vector; hence, the two equations are consistent and at least one of $p^1$ and $p^2$ is non-zero.

One must be careful to distinguish the geometry of a *reduced connection* $\varpi$ on an $\{e\}$-structure with the geometry of the manifold defined by the connection that makes that frame field parallel.  In particular, $\varpi$ is trivial since $\mathfrak{g} = 0$, but the latter connection takes its values in $\mathfrak{gl}(4)$; the space of reduced connections then consists of only zero.

However, the automorphism group of $\{e\}(M)$ is more involved, and consists of the diffeomorphisms of $M$ that take the global frame field $\{\mathbf{t}, \mathbf{n}, \mathbf{p}, \mathbf{q}\}$ to itself, which is the essence of parallel translation.  If we denote this frame by $\mathbf{e}_\mu$ and its reciprocal coframe field by $\theta^\mu$ then the infinitesimal automorphisms of $\{e\}(M)$ are the vector fields $\mathbf{v}$ on $M$ such that either:

$$0 = [\mathbf{v}, \mathbf{e}_\mu] = v^\nu[\mathbf{e}_\nu, \mathbf{e}_\mu] - \mathbf{e}_\mu v^\nu \mathbf{e}_\nu, \tag{5.52}$$

so:

$$\mathbf{e}_\mu v^\lambda + c^\lambda_{\mu\nu} v^\nu = 0, \tag{5.53}$$

or, dually:

$$0 = \mathbf{L}_\mathbf{v} \, \theta^\mu = dv^\mu + v^\nu i_{\mathbf{e}_\nu} d\theta^\mu, \tag{5.54}$$

so:

$$dv^\mu + v^\lambda c^\mu_{\lambda\nu} \theta^\nu = 0, \tag{5.55}$$

for all $\lambda$, $\mu = 0, 1, 2, 3$.  One sees that the structure function $c^\lambda_{\mu\nu}(x)$ of the frame field that $[\mathbf{e}_\mu, \mathbf{e}_\nu]$ defines plays the crucial role.  If the frame members all commute, which makes $M$ diffeomorphic to $\mathbf{R}^p \Im T^{4-p}$ for some $p$, then the only infinitesimal automorphisms must have constant values of $v^\mu$, which means they are all parallel vector fields and the Lie algebra of infinitesimal automorphisms is simply $\mathbf{R}^4$.  With a slight increase in generality, if the function $c^\lambda_{\mu\nu}$ is constant then these structure constants define a non-Abelian Lie algebra $L$ over $\mathbf{R}^4$, and a faithful representation of $L$ in $X(M)$ as parallel vector fields:

$$L \rightarrow X(M), \qquad\qquad v^\mu \infty \, v^\mu \mathbf{e}_\mu \tag{5.56}$$

However, these last two possibilities imply a high degree of symmetry to $M$, and the general case will be more involved.

**6. Discussion.**  The basic objective of this ongoing series of articles is to explore the possibility that some of the general principles of nature that physics accepts in the context of condensed matter physics could be directly applied to the particular question of the phases of the spacetime vacuum manifold by means of the geometrical intermediary of *G*-structures.

In Part I, we examined the topological obstructions to each reduction of *GL*(*M*) that were defined by the sequence (0.1) and attempted to identify them as generalizations of the topological defects that condensed matter physics associates with spontaneous symmetry breaking during phase transitions. In this part, we examined the geometry of the reduced bundles in the geometrical subsequence of (0.1) by assuming that one starts with an affine connection and gradually dissects it into smaller pieces and complementary deformation 1-forms. We also attempted to give a corresponding physical interpretation for the reduced bundles and the nature of the phase transitions.

A compelling question to ask is concerned with the fact that one of the fundamental tensor fields that was defined in this sequence of reductions has considerable physical significance, namely, the metric tensor field, which is traditionally associated with the presence of gravitation in spacetime. One necessarily wonders if any or all of the other fields that are associated with the other reductions have correspondingly profound significance. It is reasonable to speculate that the role of a volume element is most definitive in the theory of the electromagnetic interaction laws of the vacuum manifold in that phase and that the reduction to a Lorentzian metric is more related to the electromagnetic constitutive laws as they relate to the propagation of electromagnetic waves (cf. [**14-16**]). Furthermore, this should also be related to the reduction to an *SO*(2)-structure since that too seems crucial to the nature of wave motion. It is also well known by now that the theory of gravitation can just as well be formulated in terms of a global frame on spacetime [**17-19**], as in terms of a Lorentzian metric. This suggests that gravity is a sort of residual structure on spacetime that is left when there are no other motions to excite the spacetime vacuum manifold, like the non-zero ground states of quantum physics.

The subject of deformations of *G*-structures was alluded to above in the context of irreducible connections that could be decomposed into reducible ones and a deformation 1-form $\tau$ that was associated with the choice of reduction. This needs to be examined in deeper detail in the context of continuum-mechanical interpretations, especially since the basic result of [**8**] was that the modulus of the Klein-Gordon wave function was not only associated with the mass density function of the extended particle that was described by the wavefunction, but also with a dilatation of the *SO*(2)-frame field that was associated with this wave function, which, in turn, led to the equivalence of the Madelung potential function with the scalar curvature of the conformal transformation of the Minkowski metric that this dilatation defines. Hence, the deformations of the reduced bundles in the geometrical subsequence and the associated reductions of the connections on them will define the subject of the next article in this series.

Ultimately, we would like to bring the topology and geometry back together in the same discussion. The most reasonable context for this discussion seems to be the manner by which topological defects cause deformations of the geometry of continuous media, such as dislocations and disclinations in plastic media. The key to relating the one to the other is examining the integrability of the *G*-structure and how the topological obstructions to the reduction might also obstruct this integrability. This topic will be addressed in Part IV.